\documentclass[a4paper,superscriptaddress,nofootinbib,12pt]{revtex4}
\usepackage{amsfonts}
\usepackage{amsmath}
\usepackage{amsthm}
\usepackage{bbm}
\usepackage{amssymb}
\usepackage{mathrsfs}
\usepackage{color}
\usepackage[hidelinks]{hyperref}
\usepackage{stmaryrd}
\usepackage{mathrsfs}
\usepackage{dsfont}
\usepackage{wasysym}
\usepackage{eufrak}
\usepackage[a4paper]{geometry}
\usepackage{graphicx}
\def\be{\begin{equation}}
\def\ee{\end{equation}}

\def\d{\mathrm{d}}
\def\sk{\vskip .5cm}
\def\FF{\mathcal{F}}

\def\RR{\mathcal{R}}
\def\oR{{\bar{R}}}
\def\of{{\bar{\mathrm{f}}}}

\def\f{{{\mathrm{f}}}}

\def\al{\alpha}
\def\be{\beta}
\def\st{\star}
\def\epsi{{\varepsilon}}
\def\eq{\begin{equation}}
\def\la{\lambda}
\def\id{{\rm id}}
\def\en{\end{equation}}
\def\LL{{\cal L}}
\def\DD{{\cal D}}
\def\OK{{\cal{O}}\Big(\frac{1}{\kappa^2}\Big)}

\def\Ok{{\cal{O}}\big(\frac{1}{\kappa^2}\big)}
\def\om{\omega}
\def\xb{\boldsymbol{x}}
\def\kb{\boldsymbol{k}}

\def\pb{\boldsymbol{p}}

\newcommand{\D}{\Delta}
\usepackage{yfonts}
\newcommand{\pw}{\mbox{\textswab{pw}}}

\begin{document}

\title{Observables and Dispersion Relations in $\kappa$-Minkowski
  Spacetime} 
\author{Paolo Aschieri}
\email{aschieri@to.infn.it}
\affiliation{Dipartimento di Scienze e Innovazione Tecnologica,
Universit\`a del Piemonte Orientale, Viale T. Michel 11, 15121 Alessandria, Italy}
\affiliation{INFN, Sezione di Torino,  via P. Giuria 1, 10125 Torino}
\affiliation{Arnold$\,$-Regge Center, via P. Giuria 1, 10125 Torino}
\author{Andrzej Borowiec}
\email{andrzej.borowiec@ift.uni.wroc.pl}
\affiliation{Institute of Theoretical Physics, University of Wroclaw, pl. M. Borna 9,
50-204 Wroclaw, Poland.}
\author{Anna Pacho{\l }}
\email{a.pachol@qmul.ac.uk}
\affiliation{Queen Mary, University of London, School of
  Mathematics, Mile End Rd.,
London E1 4NS, UK.}

\begin{abstract}
{\vskip .2cm}
\noindent
We revisit the notion of quantum Lie algebra of symmetries of a
noncommutative spacetime, its elements are shown to be the generators
of infinitesimal transformations and are naturally identified with
physical observables. Wave
equations on noncommutative spaces are derived from a quantum Hodge 
star operator. 

This general noncommutative geometry construction is then exemplified
in the case of $\kappa$-Minkowski spacetime. The corresponding quantum Poincar\'e-Weyl
Lie algebra of infinitesimal translations, rotations and dilatations
is obtained. The d'Alembert wave operator
coincides with the quadratic Casimir of  quantum translations and it is
deformed as in Deformed Special Relativity theories. Also momenta
(infinitesimal quantum translations) are deformed, and
correspondingly the Einstein-Planck relation and the de Broglie one.
The energy-momentum relations (dispersion relations) 
are consequently deduced. These results complement those of the
phenomenological literature on the subject.
\end{abstract}
\maketitle
\noindent {\small Keywords: Noncommutative Geometry, Twist deformation,
  Quantum Lie algebra, \\ $~~~~~~~~~~~~~~~$ Dispersion Relations,
  $\kappa$-Minkowski spacetime,  DSR theories}

\tableofcontents

\section{Introduction}
The open challenge of a quantum description of the gravitational
interaction and the corresponding notion of quantum spacetime 
is particularly demanding because of the very limited
experimental hints at our disposal. Indeed 
the quantum structure of spacetime, due to quantum gravity effects, is expected to
become relevant at Planck scales ($l_p=\sqrt{\frac{G}{\hbar c^3}}\sim
10^{-33}$ cm).  Since these energy regimes are
not directly accessible we have to study indirect signatures, for example
related to cosmological data near inflationary epoch (a few
orders of magnitude away from regimes characterised by Planck scale
energies) or related to modified dispersion relations of light.

Indeed it is conceivable that light travelling in a quantum spacetime
(a dynamical spacetime that interacts with photons) has a velocity
dependent on the photons energy. 
Even a tiny modification of the usual dispersion relations could then
be detected due to the cumulative effect of light travelling long
distances.
A natural setting for this study is that of Gamma Ray Bursts (GRB)
from distant galaxies \cite{AmelinoCamelia:1997gz}. Another possibility is that of
high precision (quantum) optics experiments based on interferometry
techniques like the Holometer one in
Fermilab \cite{Hogan:2010zs} or the table top experiment devised in \cite{Genovese}.

These possibilities motivate the study of phenomenological
models that lead to modified dispersion relations. Lorentz invariance
violating (LIV) theories generically provide modified dispersion
relations, see for example \cite{Liberati}.
A whole class where the Lorentz group (or its realization) is modified,
so that new relativistic symmetries replace the classical one, goes
under the name of Deformed (or Doubly) special relativity theories
(DSR theories) \cite{AmelinoCamelia:2000mn, MS1, MS2}. Many of these
phenomenological models describe spacetime features and wave equations that are typical of noncommutative
spacetimes, the prototypical example being $\kappa$-Minkowski
spacetime, where coordinates satisfy the relations
$  x^{0}\st x^{\/j}-x^{\/j}\st x^0=\frac{i}{\kappa }x^{\/j}\,,~x^{i}\st x^{j}-x^j\st x^i=0$; 
here we consider $\frac{1}{\kappa}$ of the order of the Planck
length. See \cite{KowalskiGlikman:2001gp} for an early relation between
DSR and  $\kappa$-deformed symmetries.
The relation between DSR theories and noncommutative spacetimes is very interesting because noncommutative spacetimes,
and their quantum symmetry groups, 
independently from DSR studies, naturally arise
as models of quantum spacetime where discretization and
indetermination relations are realized in spacetime itself rather
than phase-space. This is indeed expected by gedanken
experiments probing spacetime structure at Planck scale;
furthermore noncommutative spacetimes features, like generalized uncertainty relations (involving space and time coordinates) or minimal area and volume elements arise also in String Theory and Loop Quantum Gravity, see e.g. \cite{Hossenfelder}.
\sk
\sk

In this paper we complement the
phenomenological or bottom up approach of constructing (realizations
of) spacetime symmetries, observables and field equations that model
expected and possibly detectable quantum gravity effects 
with a top-down approach where we just assume a general noncommutative
structure of spacetime and then use noncommutative differential
geometry to derive the physics of fields propagating in this
noncommutative spacetime. 

We focus on spacetimes that are obtained via a Drinfeld twist procedure, these
form a very large class of noncommutative spacetimes (including the
most studied Moyal-Weyl one $x^i\star x^j-x^j\star x^i=i\theta^{ij}$
\cite{Chaichian:2004za} and the $\kappa$-Minkowski
one, $x^0\star x^i-x^i\star x^0=\frac{i}{\kappa}x^i$ \cite{ABTwist, Borowiec:2008uj}). 
We present a general method on how to study observables, wave
equations and the corresponding dispersion relations,
thus providing a wide framework for physical  models of  fields in
noncommutative spaces.
The construction is quite constrained by the mathematical
consistency of the noncommutative differential geometry and the
notion of infinitesimal symmetry generators
(e.g. translations-momenta).
We thus contribute filling the gap between theory and
phenomenology, providing sound and mathematically consistent tools for
phenomenological studies as well as a deeper physical understanding of
the mathematical structures associated with noncommutative spaces.

We then apply the theory thus developed to the $\kappa$-Minkowski
spacetime and derive the same equations of
motion studied in the phenomenological (bottom up) approach considered
in \cite{MS1}. 
 
We show that the usual Einstein-Planck relation $E=\hbar\omega$ is modified to
$\hbar\omega/(1-\omega/c\kappa)$ leading to a maximum Planck energy or
frequency per elementary particle. Similarly, the de Broglie momentum-wavelength relation is modified. This implies 
that massless fields in $\kappa$-Minkowski spacetime have no modified
dispersion relations, in agreement with \cite{MS1} and disagreement with \cite{BGMP, AmelinoCamelia:2011cv}. 
%
%
%
%
  
 As we show in the sequel to this paper \cite{ABP2}
the methods here developed do in general imply modified dispersion
relations for massless fields once we leave flat noncommutative
spacetime by turning on a nontrivial metric background. 
\sk
\sk
Our primary interest in the present paper is the equation of motion of
massless fields in flat spacetime. In the commutative case these
propagate in Minkowski spacetime, and the relevant symmetry group is the
conformal group.  Hence we consider the noncommutative
differential geometry on $\kappa$-Minkowski spacetime that is
covariant under the action of a quantum conformal group (in particular
the Minkowski metric won't be invariant but covariant under quantum dilatations). For
simplicity and brevity rather than deforming the conformal group we
focus on the most relevant part, the Poincar\'e-Weyl (also simply
called Weyl) group of Poincar\'e transformations and dilatations, the
inclusion of the special conformal transformations being
straightforward. The choice to focus on  Poincar\'e-Weyl symmetry
rather than Poincar\'e symmetry is motivated by the study of massless
fields and the interest in applying the general noncommutative
differential geometry construction via twist deformation that we develop in \S \ref{II} to
the case of the well known $\kappa$-Minkowski noncommutative
spacetime. 

We further observe that the only term breaking conformal symmetry in the standard model
action is the Higgs mass term and that there are interesting scenarios
where there is no such term in the classical action, so that the conformal group is a
fundamental symmetry of the tree level action. There, according to
the seminal work \cite{CW},  the
standard model masses are generated via radiative corrections that
break the classical scale invariance of the Higgs potential thus inducing the spontaneous symmetry breaking of the electroweak interaction.
These models, see e.g. \cite{Englert}, provide a solution to the hierarchy problem
up to Planck scale thanks to classical scale invariance. It
is then intriguing to
speculate that due to quantum gravity effects the conformal symmetry
at Planck scale is twist deformed to the quantum conformal symmetry
associated with the quantum Poincar\'e-Weyl Lie algebra that we study in
\S \ref{IIIA1}. The deformation parameter  $\frac{1}{\kappa}$ is
indeed dimensionful and considered of the order of the Planck length.

A key point in order to  have a physical interpretation of the wave
equation (e.g. of the corresponding dispersion relations) is that of 
identifying the physical generators of Poincar\'e-Weyl transformations
and in particular the four-momentum operators. 
In this paper we first attack the problem mathematically and show
that given a twist (hence a given quantum group, deformation of a
usual group) there is a unique notion of quantum Lie algebra (with
quantum Lie bracket and quantum Jacobi identity). Quantum Lie algebra elements act on
fields on noncommutative spacetime as quantum infinitesimal
transformations because they satisfy a specific deformation of the
Leibniz rule. They are obtained via a quantization map $\cal D$
applied to classical Lie algebra elements. Then we confirm their interpretation as generators
of infinitesimal transformations by recalling their associated differential
geometry; in particular on $\kappa$-Minkowski space the quantum Lie
algebra of momentum generators is shown to give the
differential calculus.
Because of these properties these quantum Lie algebra generators are
naturally identified with the physical observables. Of course it would
be interesting to confirm this identification with a Noether theorem
analysis. Noether theorem for theories on noncommutative spacetimes
has to be further understood. For preliminary work we refer for example to~\cite{ACAC}. The methods we
advocate, possibly combined with  Drinfeld twist deformed Hamiltonian
mechanics, see e.g.~\cite{ALV},
should provide powerful tools in the study of conserved charges in noncommutative spacetimes.

The same quantization map $\cal D$ used to single out the quantum Lie algebra of
infinitesimal generators from the undeformed Lie algebra
is then used to obtain the quantum Hodge star operator from the usual
Hodge star operator. This procedure is very general and does not need to
rely on a flat metric. The associated Laplace-Beltrami
operator and the corresponding wave equation is then presented. In the
case of $\kappa$-Minkowski space the Laplace-Beltrami
operator becomes the d'Alembert operator and we show that it
equals the quadratic Casimir of quantum translations.
This implies covariance of the wave equation under the
quantum Poincar\'e-Weyl group. Hence, in particular, besides the parameter $c$ (velocity of
light) also  the dimensionful deformation parameter
$\frac{1}{\kappa}$ (the Planck length) is constant under quantum
Poincar\'e-Weyl transformations. In other words, as usual in
noncommutative geometry, the quantum group generalization of the
notion of symmetry allows to extend the principle of relativity
of  frames to that of relativity of  frames related by quantum symmetry 
transformations\footnote{All frames related by quantum Poincar\'e-Weyl transformations detect
the same values of $c$ and $\frac{1}{\kappa}$. In this paper for
simplicity we do not consider finite transformations, we focus on
quantum Lie algebras and infinitesimal transformations, with 
$c$ and $\frac{1}{\kappa}$ that are constant under these transformations.}, (cf. also \cite[\S 8]{book}).
\sk
\sk
Finally, we compare our results with the pioneering approach in
\cite{MS1, MS2}.
We show that the nonlinear realization of the Lorentz group considered in
\cite{MS1} is implemented by the quantization map ${\cal D}$
that gives the Poincar\'e-Weyl quantum Lie algebra. 
Since the quantization map ${\cal D}$ does not depend on the 
representation considered, this allows to extend
 the construction in \cite{MS1}, based on momentum space, to
arbitrary representations; in particular to position space.
Furthermore, in our scheme the wave equation turns out to be the same as that in
\cite{MS1}, however the identification of the physical momenta
differs  from that in \cite{MS1}.

\section{Twist Noncommutative Geometry}\label{II}

The noncommutative deformations of spacetime we consider arise from
the action of a symmetry group (group of transformations) on
spacetime, e.g. Poincar\'e, Poincar\'e-Weyl, or conformal groups
 on Minkowski spacetime.  
Physical observables on these noncommutative spacetimes are
related to their symmetries. Like in the commutative case, where
energy and momentum
are the generator of spacetime translation, in the noncommutative case we identify energy and
momentum as the generators of noncommutative spacetime translations. 
Since infinitesimal transformations are obtained as actions of the Lie
algebra of the symmetry group on the spacetime and on the matter
fields, we first have to recall the notion of quantum group and
quantum Lie algebra of a quantum group. Then we consider their symmetry
transformations on the corresponding noncommutative spaces. The
construction of the differential geometry on these spaces, covariant
under these symmetry transformations, leads to wave equations
describing fields propagating in noncommutative spacetimes.

\subsection{Quantum Symmetries and  Quantum Lie algebras}\label{IIA}

Let $G$ be a Lie group of transformations on a manifold $M$; infinitesimal
transformations are governed by its Lie algebra $g$, or equivalently
by the universal enveloping algebra $Ug$. We recall that $Ug$ is the
sum of products of elements of $g$ (modulo the Lie algebra relations)
and that it is a Hopf algebra with coproduct 
$ \Delta:Ug\rightarrow Ug\otimes Ug$, counit $\epsilon : Ug\rightarrow \mathbb{C}$
and antipode $S$ given on the generators $u\in g$ as:
$\D(u)=u \otimes 1 + 1 \otimes u,\, \epsi (u)=0, \, S(u)=-u$
and extended to all $Ug$ by requiring 
$\D$ and $\epsi$ to be linear and multiplicative
while $S$ is linear and antimultiplicative.
In the following we use Sweedler notation for the coproduct
\eq
\Delta(\xi)=\xi_1\otimes \xi_2
\en
where $\xi\in Ug$,  $\xi_1\otimes \xi_2\in Ug\otimes Ug$ and a sum
over $\xi_1$ and $\xi_2$  is understood ($\xi_1\otimes
\xi_2=\sum_i\xi_{1i}\otimes \xi_{2i}$).
There is a canonical action of $Ug$ on itself obtained from the
coproduct and the antipode, it is given by the
adjoint action, defined by, for all $\zeta,\xi\in Ug$, 
\begin{equation}
\zeta(\xi)\equiv ad_\zeta(\xi)=\zeta_1\xi S(\zeta_2)~.\label{adact}
\end{equation}
When restricted to Lie algebra elements this becomes the Lie bracket,
indeed we have, for all $u,u'\in g$,
\eq
[u,u^{\prime }]=u_1 u^\prime
S(u_2)=u u'-u^\prime u\label{resadact}
\en 
where we used that
$\Delta(u)=u\otimes 1+1\otimes u$ and that $S(u)=-u$.
Notice that the $Ug$-adjoint action restricts to and action on $g$,
i.e., for all $\xi\in Ug$, $v\in g$, $\zeta(v)\in g$. This is easily
seen by writing $\zeta$ as sums of products $uu'\ldots u''$ of Lie
algebra elements, and
then by iteratively using (\ref{resadact}): $(uu'\ldots u'')(v)=[u[u'...[u'',v]]]$.
\sk
We deform spacetime by first deforming the Hopf algebra $Ug$ and then
functions and matter fields on spacetime itself.
Since we work in the deformation quantization context we extend the notion of enveloping algebra to formal power series in $\lambda$ and we
correspondingly consider the Hopf algebra $Ug[[\la]]$ over the ring
$\mathbb{C}[[\la]]$ of formal power series with complex coefficients. In
the following, for the sake of brevity we will often denote $Ug[[\la]]$
by $Ug$.\footnote{Physically we are interested in the lowest order
  corrections in $\la$ to dispersion relations and related
  expressions. In this case $\la$ is linked to the Planck energy $E_p$ and the expansion will be in  $\hbar\om/{E_p}$ with $\om$ the
frequency of the wave. The lowest order corrections are then well
defined also nonformally, i.e. with $\hbar\om/{E_p}$ a real
number. Specific expression can be well defined nonformally also at
all powers in $\hbar\om/{E_p}$, see e.g. the Einstein-Planck relations
(\ref{eigvlEp}). The treatment of $\la$ as a formal parameter is
a mathematical requirement for the $\star$-product between arbitrary
smooth functions to be well defined.  However for the exponential (hence analytic)
functions we consider as solutions of the wave equation
it is consistent to consider $\hbar\om/{E_p}$  a real number.}

\sk
Following Drinfeld, a twist element $\mathcal{F}$ on the universal enveloping algebra $Ug[[\la]]$ of a
Lie algebra $g$ is an invertible element of $Ug[[\la]] \otimes
Ug[[\la]] $ satisfying the cocycle and normalisation conditions: 
\begin{equation}
(\mathcal{F}\otimes 1)(\Delta \otimes id)\mathcal{F}=(1\otimes \mathcal{F}%
)(id\otimes \Delta )\mathcal{F},  \label{Twcond1}
\end{equation}%
\begin{equation}
  (id\otimes \epsilon )\mathcal{F}=(\epsilon \otimes id)\mathcal{F}=1\otimes 1~.
\label{Twcond2}
\end{equation}%
The first relation implies associativity of the $\star$-products induced
by the twist; consistently with the second relation we also require that $\FF=1\otimes 1+ {\cal
  O}(\lambda)$ so that for $\la\to 0$ we recover the classical
(undeformed) products.
\sk
Given a twist $\mathcal{F}$ on the Hopf algebra $Ug$ we have a new Hopf
algebra structure $Ug^{\mathcal{F}}=\left( Ug,m, \Delta^{\mathcal{F}},\epsilon ,S^{\mathcal{F}}\right) $;
$Ug^\FF$ equals $Ug$ as  vector space and also as an algebra (the
product $m$ is undeformed, and also the counit $\epsi$) the deformed
coproduct and antipode are defined by, for all $\xi \in Ug $, 
\begin{equation}
\Delta ^{\mathcal{F}}(\xi ) =\mathcal{F}\Delta (\xi )\mathcal{F}^{-1}\quad
,\quad\quad S^{\mathcal{F}}(\xi )=\chi
S(\xi )\chi^{-1} \label{TwistedUg}
\end{equation}%
where $\chi=\mathrm{f}^{\alpha  
}S(\mathrm{f}_{\alpha })$ and we used the notation: $\mathcal{F}=\mathrm{f}^{\alpha }\otimes \mathrm{
f}_{\alpha },\quad \mathcal{F}^{-1}=\bar{\mathrm{f}}^{\alpha }\otimes \bar{%
\mathrm{f}}_{\alpha }~$ (sum over $\alpha$ understood). 
We denote by $ad^\FF: Ug^\FF\otimes Ug^\FF\to Ug^\FF$ the
$Ug^\FF$-adjoint action, it is given by, for all $\xi, \zeta\in
Ug^\FF$,  $ad^\FF_\xi(\zeta)=\xi_{1^\FF}\zeta S^\FF(\xi_{2^\FF})$,
where we used the Sweedler notation $\Delta^\FF(\xi)=\xi_{1^\FF}\otimes\xi_{2^\FF}$.
\sk
We now study the quantum Lie algebra $g^\FF$ 
of the Hopf algebra $Ug^\FF$. To this aim let us recall that
there is a one-to-one correspondence between Lie algebras $g$ and 
universal enveloping algebras $Ug$: given $g$ then $Ug$ is canonically
constructed, vice versa $g\subset Ug$ is the
subspace of primitive elements of $Ug$, i.e., of those elements that satisfy
$\Delta(u)=u\otimes 1+1\otimes u$. From this expression it is not
difficult to show that $\epsi(u)=0$ and $S(u)=-u$ so that as in
(\ref{resadact}) the adjoint action of two primitive elements $u$ and
$v$ equals their commutator:  $[u,v]:=ad_u(v)=uv-vu$.
Now it is easy to show that $uv-vu$ is again a primitive element 
thus proving that $g$ with the bracket $[~,~]$ is a Lie algebra: the
Lie algebra of the universal enveloping algebra $Ug$.
Similarly, in the quantum case, following \cite{Woronowicz},  we have
a quantum Lie algebra $g^\FF$ of a quantum universal enveloping
algebra $Ug^\FF$ when
\begin{itemize}

\item[{\it i)}] $g^\FF$ generates $Ug^\FF$,  

\item[{\it ii)}] $\Delta^\FF(g^\FF)\subset g^\FF\otimes 1 + Ug^\FF\otimes g^\FF$

\item[{\it iii)}] $[g^\FF,g^\FF]_\FF\subset g^\FF$
\end{itemize}
where the quantum Lie bracket $[~,~]_\FF$ is the restriction of the
$Ug^\FF$-adjoint action  to $g^\FF$, i.e., for all $u^\FF,v^\FF\in g^\FF$ we have
\begin{equation}
[u^\FF,v^\FF]_\FF:=u_{~{1^\FF\,}}^\FF{} v^\FF S^\FF(u_{~2^\FF}^\FF)~.\label{adjact[]}
\end{equation} 
Property {\it iii)} states
that the restriction to $g^\FF$ of the $Ug^\FF$-adjoint action is well
defined on $g^\FF$. Since from Property {\it i)} we have that $g^\FF$
generates $Ug^\FF$ we conclude that $g^\FF\subset Ug^\FF$ is invariant
under the $Ug^\FF$-adjoint action, i.e. for all $\xi\in Ug^\FF$, $v^\FF\in g^\FF$,
$ad^\FF_{\xi}(v^\FF)\in g^\FF$. For example $ad^\FF_{u^\FF  u'^\FF}(v^\FF)=
ad^\FF_{u^\FF}(ad^\FF_{u'^\FF}(v))=[u^\FF,[u'^\FF,v]_\FF]_\FF\in g^\FF$.

Property {\it ii)}  states that the elements in $g^\FF$ are
quasi-primitive, i.e., we require a minimal deviation from the usual
coproduct property $\Delta(u)=u\otimes 1+ 1\otimes u$ of elements
$u\in g$. Combining  {\it ii)}  with {\it iii)} it can be shown
\cite{Aschieri:1995wg} \cite[\S 2.3]{Tesi} that the bracket $[~,~]_\FF$ is
a deformed commutator which  is quadratic in the generators of the Lie algebra $g^\FF$.
\sk

In the twist deformation case we are considering there is a canonical
construction in order to obtain the quantum Lie algebra $g^\FF$ of
the Hopf algebra $Ug^\FF$ \cite{GR2} \cite[\S 7.7]{book}. We simply have $g^\FF=\DD(g)$, i.e., 
\begin{equation}
g^\mathcal{F}:=\{u^\FF\in Ug^\FF; \; u^\FF\!=\DD(u) , \mbox{ with } u\in g\} \label{twistedgen}
\end{equation}
where, for all $\xi\in Ug$,
\eq\label{DonUg}
\DD(\xi)={\bar{\mathrm{f}}}^\alpha(\xi){\bar{\mathrm{f}}}_\alpha=\of^\al_1\xi
S(\of^\al_2)\of_\al~,
\en 
the action  $\of^\alpha(\xi)$ being the adjoint action (\ref{adact}).  
Property {\it i)} follows form the twist property $\FF=1\otimes 1+
{\cal O}(\la)$ that implies $\DD=\id+ {\cal O}(\la)$.\footnote{
{\it Proof.} Since as vector
spaces $Ug^\FF=Ug[[\la]]$, any element $\xi\in Ug^\FF=Ug[[\la]]$ is a sum
$\xi=\sum_{j\geq 0}\la^j\xi_j$ with  $\xi_j\in Ug$ (and no $\la$
dependence in $\xi_j$). Since $g$ generates $Ug$ we have
$\xi_j=\sum_{m_1...m_n}c_{j, m_1,...m_n}u_1^{m_1}u_2^{m_2}...u_n^{m_n}$ where
$u_1, u_2,...u_n$ are a basis of $g$. Then we have
$$\xi=\sum_j\la^j\xi_j=\sum_{j,m_1,..m_n}
\la^jc_{j,m_1,...m_n}u_1^{m_1}...u_n^{m_n}=
\sum_{j,m_1,..m_n}
\la^jc_{j,m_1,...m_n}\DD(u_1)^{m_1}...\DD(u_n)^{m_n}\,+\,\la\xi^{(1)}
$$
where $\xi^{(1)}\in Ug[[\la]]$. We can therefore expand  
$\xi^{(1)}$ in the same way as done with $\xi$ and obtain
$\la \xi^{(1)}=\sum_{j,m_1,..m_n}
\la^{j+1}c^{(1)}_{j,m_1,...m_n}\DD(u_1)^{m_1}\ldots\DD(u_n)^{m_n}+\la^2\xi^{(2)}
$. Iterating this procedure we thus arrive at the expression
$\xi=
\sum_{j,m_1,..m_n}
(\la^jc_{j,m_1,...m_n}
+\la^{j+1}c^{(1)}_{j,m_1,...m_n}
+\ldots \la^{j+k}c^{(k)}_{j,m_1,...m_n})
\DD(u_1)^{m_1}\ldots\DD(u_n)^{m_n}+ \la^{(k+1)}\xi^{(k+1)}$
that shows that $\xi$ is generated by the elements $\DD(u_1), \DD(u_2),
\ldots \DD(u_n)$ because this is so up to order $\la^k$ with $k$ an
arbitrarily big integer $\boxempty.\;$
{}An explicit proof for the algebra we will consider 
follows immediately from 
(\ref{FF40}) and (\ref{tildeexp}).}
Property {\it ii)} holds because, cf.  (3.12) and (3.17) in \cite{GR2},
for all $u\in g$,
\eq\label{eleven}
\Delta^\FF (u^\FF)=u^\FF\otimes 1+\oR^\al\otimes (\oR_\al(u))^\FF
\en
where $\mathcal{R}=\mathcal{F}_{21}\mathcal{F}^{-1}$ is the so-called
universal $R$-matrix, and we used the notation $\mathcal{R}^{-1}={\bar{R}}^\alpha\otimes{\bar{R}}%
_\alpha$ (the action  ${\bar{R}}^\alpha(u)$ of  ${\bar{R}}^\alpha$ on
$u\in g$ is the adjoint action (\ref{adact}), and
${\bar{R}}^\alpha(u)\in g$, cf. sentence after (\ref{resadact})). 
Property {\it iii)}  holds because it is equivalent
to equation (4.6) in \cite{GR2} (just apply $\DD$ to (4.6)).  Similarly
to the undeformed case we then have that the adjoint action (\ref{adjact[]})
equals  the braided commutator, for all $u^\FF, v^\FF\in g^\FF$, 
\begin{equation}  \label{TwistedCom}
[u^\FF,v^\FF]_\mathcal{F}=u^\FF v^\FF-({\bar{R}}^\alpha(v))^\FF_{\,}
 ({\bar{R}}_\alpha(u))^\FF~.
\end{equation}
Furthermore the deformed Lie bracket satisfies the
braided-antisymmetry property and the braided-Jacobi identity
\begin{eqnarray}\label{braid1}
&&[u^\FF,v^\FF]_\mathcal{F} =-[({\bar{R}}^\alpha(v))^\FF, ({%
\bar{R}}_\alpha(u))^\FF]_\mathcal{F} \\[.3em]
&&[u^\FF ,[v^\FF,z^\FF]_\FF ]_\mathcal{F} =[[u^\FF,v^\FF]_\FF ,z^\FF]_\mathcal{F} + [({\bar{R}}%
^\alpha(v))^\FF, [({\bar{R}}_\alpha(u))^\FF,z^\FF]_\mathcal{F} ]_\mathcal{F} ~.\label{braid2}
\end{eqnarray}
A proof of (\ref{TwistedCom})-(\ref{braid2})  
follows immediately by applying $\DD$ to eq. (3.7) (3.9), (3.10) in
\cite{GR2}; indeed the quantum Lie algebras presented here and there
are isomorphic via $\DD^{-1}$ (cf. also \cite[\S 7 and \S 8]{book}).
\sk
In the classical case the physical operators associated with a symmetry Hopf
algebra $Ug$ are the elements of the Lie algebra $g$ (e.g. momenta boosts
and rotations in the universal enveloping algebra of the Poincar\'e
group). 
It is then natural that the physical operators associated with a symmetry Hopf
algebra $Ug^\mathcal{F}$ are the elements of the quantum Lie algebra 
$g^\mathcal{F}$. To clarify this point we have to study representations of $Ug^\FF$.

\subsection{Noncommutative algebras}
Let  $G$ be a Lie group of transformations on a manifold $M$
(e.g. the spacetime manifold) and let $A$ be the algebra of smooth
functions on $M$  (extended to formal power series in
$\la$). If the action of
$G$ on $M$ is via diffeomorphisms, then the
Lie algebra $g$ acts on $A$ via the Lie derivative, 
which is extended as an action to all $Ug$, i.e., for all $\xi,\zeta\in Ug$
$\LL_\xi\LL_\zeta =\LL_{\xi\zeta}$. The compatibility between the $Ug$
action and the product of $A$ is encoded in the property
that the Lie algebra
elements $u\in g$ act as derivations, i.e., for all $u\in g$ and $f,h,\in A$, 
$\LL_u(fh)=\LL_u(f)h+f \LL_u(h)$, or simply $u(fh)=u(f)h+f u(h)$. 
Since $u(fh)=u(f)h+f u(h)=u_1(f)u_2(h)$ and the $u$'s generate $Ug$
this implies that
 for all $\xi\in Ug$, $f,h\in A$, 
\eq\label{ugmodxfh}
\xi(f h)=\xi_1(f) \,\xi_{2}(h)~.
\en 
We say that $A$ is a $Ug$-module
algebra because it carries a representation of $Ug$ and the
$Ug$-action is compatible with the product of $A$ as in (\ref{ugmodxfh}).

Corresponding to the deformation $Ug\to Ug^\mathcal{F}$ we have the
algebra deformation $A\to A_\st$. 
As vector spaces $A$ and $A_\st$ coincide. The action of $Ug^\FF$ on
$A^\FF$ is also the same as that of $Ug$ on $A$. The space
$A_\st$ is therefore still the space of smooth functions on $M$; it is
the product that is deformed in the $\st$-product,
\eq\label{starprodfunc}
f\st h=\mu \{\mathcal{F}^{-1}(f\otimes h)\}=\of^\al(f)\,\of_\al(h)~,
\en
for all functions $f, h\in A_\st$, where $\of^\al(f)=\LL_{\of^\al} (f)$.
Associativity of the product follows from the twist cocycle property
(\ref{Twcond1}), moreover the product has been defined so that it is 
compatible with the $Ug^\FF$ action: for all
$\xi\in Ug^\FF$, $\xi(f\st h)=\xi_{1^\FF}(f) \,\xi_{2^\FF}(h)$. This
proves that  $A_\st$ is a $Ug^\FF$-module algebra.
Equivalently we say that it is a $Ug^\FF$-module algebra because the quantum Lie
algebra elements $\DD(u)\in g^\FF$ act as twisted derivations, i.e.,
they obey
the deformed Leibniz rule (cf. (\ref{eleven})),
\eq\label{LeibFF}
\DD(u)(f\st h)=\DD(u)(f)\st h+\oR^\al(f)\st \DD(\oR_\al(u))(h)~.
\en

A special case is when the manifold $M$ is the group manifold $G$ itself, then the
Lie algebra $g$ is identified with that of left invariant vector
fields on $G$. The differential geometry
on $A$ (smooth functions on $G$) is described via these
vector fields and the corresponding left invariant one-forms. Similarly,
the differential geometry on $A_\st$ (the quantum group)
is described in terms of the quantum Lie algebra $g^\FF$: the
quantum Lie algebra of left invariant vector fields on $A_\st$
\cite{Woronowicz, AC}.
\sk

We can now sharpen our claim:
among the many generators of the universal enveloping algebra $Ug^\FF$
the physically relevant ones are the elements of the  quantum Lie algebra
$g^\FF$ (and not for example of $g$). This is so because
of their geometric properties: they generate $Ug^\FF$, they are closed under the $Ug^\FF$-adjoint action and their action on representations is via twisted
derivations, so that they are {\sl quantum infinitesimal transformations}.
\vskip .4cm

\subsection{Differential calculus}\label{DiffCalc}
Consider the algebra $A$ of smooth functions
on spacetime $M$ and the action of the Lie algebra $g$ on $A$ via the
Lie derivative. The Lie derivative lifts to the algebra
$\Omega^\bullet=A\oplus\Omega^1\oplus\Omega^2\oplus...$ of exterior forms on $M$. We then twist deform this
algebra to the algebra $\Omega^\bullet_\st$, which as a vector space
is the same as $\Omega^\bullet$ (more precisely we should write
$\Omega^\bullet[[\la]]$, i.e. power series in
the deformation parameter $\la$ of classical exterior forms)  
but has the new product 
\eq
\omega\wedge_\st\omega'=\of^\al(\omega)\wedge\of_\al(\omega')~.
\en
In particular when $\omega$ is a zero form $f$, then the wedge product
is usually omitted and correspondingly the wedge $\st$-product
reads $f\st \omega'=\of^\al(f)\of_\al(\omega')$.

Since the action of the Lie algebra $g$ on forms is via the Lie derivative
and the Lie derivative commutes with the exterior derivative the usual
(undeformed) exterior derivative satisfies the Leibniz rule
$
\mathrm{d}(f\star g) =\mathrm{d}f\star g+f\star \mathrm{d}g,  
$
and more in general, for forms of homogeneous degree $\omega\in \Omega^r$, 
\eq
\d(\omega\wedge_\st\omega')=\d\omega\wedge_\st\omega'+(-1)^r\omega\wedge_\st\omega'~.
\en
We have constructed a differential calculus on the deformed algebra of
exterior forms $\Omega_\st^\bullet$.

\subsection{Hodge star operator}
Another key ingredient in order to formulate field theories on a
spacetime $M$ is a metric. It acts on exterior forms via the Hodge star
operator.
Let us first recall few facts related with the Hodge star in the classical
(undeformed) case.
For an $n-$dimensional manifold $M$ with metric $g$ 
the Hodge $\ast $-operation is a linear
map $\ast :\Omega ^{r}\left( M\right) \rightarrow \Omega ^{n-r}\left(
M\right)$. In local coordinates an $r$-form is given by
$\omega=\frac{1}{r!}\omega _{\mu _{1....}\mu_{r}}\d x^{\mu_1}\wedge\ldots
\d x^{\mu_r}$  and the Hodge $\ast$-operator reads
\eq\ast \omega =\frac{\sqrt{g}}{r!\left( n-r\right) !}\omega _{\mu _{1....}\mu
_{r}}\epsilon _{~\ \ \ \ \ \ \ \ \ \ \!\!\!\!\!\!\!\!\!\nu _{r+1}......\nu _{n}}^{\mu
_{1....}\mu _{r\,}}\d x^{\nu _{r+1}}\wedge \ldots \d x^{\nu _{n}}~
\en
where $\sqrt{g}$ is the square root of the absolute value of the determinant of the metric, the completely antisymmetric tensor $\epsilon_{\nu_1\ldots
  \nu_n}$ is normalized to $\epsilon_{1\ldots n}=1$ and indices are
lowered and raised with the metric $g$ and its inverse.
There is a one to one correspondence between metrics and Hodge star
operators. 

We define metrics on noncommutative spaces by defining the
corresponding Hodge star operators on the $\st$-algebra of exterior forms
$\Omega_\st^\bullet$. We first observe that the undeformed Hodge
$\ast$-operator is $A$-linear:
$\ast(\omega f)=\ast(\omega)\,\! f$, for any form $\omega$ and function
$f$
(of course, since $A$ is commutative we equivalently  have
$\ast(f\omega)=f\,\!(\ast\omega)$). 
We then require the Hodge $\ast$-operator $\ast^\FF$ on $\Omega_\st^\bullet$ to
map $r$-forms into $n-r$-forms, and to be right $A_\st$-linear
\eq
\ast^\FF(\omega\star f)=\ast^\FF(\omega)\star f
\en
for any form $\omega$ and function $f$.
There is a canonical way to deform $A$-linear maps to 
right $A_\st$-linear maps, this is given by the  
``quantization map'' $\DD$ studied in \cite{KM, AS}.
Let $V$ be a right $A$-module (i.e., for all $v\in V$, $f,h\in A$ we have
$vf\in V$ and $(vf)h=v(fh)$), then $V$ is a right
$A_\st$-module by defining $v\star f=\of^\al(v)\of_\al (f)$ (the
property $(v\star f)\star h=v\star (f\st h)$ follows from the cocycle
condition (\ref{Twcond1}) for the twist). Now let $V,W$ be right $A$-modules and $P:
V\to W$ be a right $A$-linear map, i.e., $P(vf)=P(v)f$, for all $v\in
V$, $f\in A$. Then, similarly to the way we obtained the quantum Lie
algebra generators we have the quantization $P^\FF=\DD(P)$ of the right 
$A$-linear map $P$,
\begin{equation}
P^\FF=\mathcal{D}\left( P\right) :
=\LL_{\mathrm{\bar{f}}_{1 }^{\alpha }}\circ P\circ \LL_{S\left( \mathrm{%
\bar{f}}_{ 2 }^{\alpha }\right) \mathrm{\bar{f}}%
_{\alpha }}
\label{quantum_map}
\end{equation}%
that is a right $A_\st$-linear map: $P^\FF(v\st f)=P^\FF(v)\st f$. 
In particular the deformed or quantum Hodge $\ast$-operator explicitly
reads
 \begin{eqnarray}\label{quantum_ast}
\ast ^{\mathcal{F}}=\mathcal{D}(\ast): \,\Omega_\st^\bullet&\longrightarrow& \Omega^\bullet_\st\nonumber\\
\omega & \longmapsto & 
\ast ^{\mathcal{F}}(\omega)={\mathrm{\bar{f}}_{1 }^{\alpha }}\Big(\!\ast\big(S({\mathrm{\bar{f}}_{ 2 }^{\alpha}})\of_\al(\omega)\big)\!\Big) ~
\end{eqnarray}
where ${\mathrm{\bar{f}}_{1 }^{\alpha
  }}\Big(\!\ast\big(S({\mathrm{\bar{f}}_{ 2
  }^{\alpha}})\of_\al(\omega)\big)\!\Big) $ is a shorthand notation
for ${\LL_{\mathrm{\bar{f}}_{1 }^{\alpha }}}\Big(\!\ast\big(\LL_{S({\mathrm{\bar{f}}_{ 2 }^{\alpha}})\of_\al}(\omega)\big)\!\Big) $.
For any exterior form $\omega$ and function $f$ we have the right
$A_\st$-linearity property $\ast ^{\mathcal{F}}(\omega\st f)=\ast
^{\mathcal{F}}(\omega)\st f$. 

Finally we notice that the metric structure we have
introduced via the Hodge star operator is  a priori independent from the
twist $\FF$ determining the noncommutativity of spacetime.

\subsection{Wave equations}
The wave equation in (curved) 
spacetime is governed by the Laplace-Beltrami operator $\Box = \delta
d + d\delta$.
In the case of even dimensional Lorenzian
manifolds (like Minkowski spacetime) the adjoint of the exterior derivative is
defined by $\delta= \ast d \ast$.
 For a scalar field (i.e., a function or 0-form) we have
\begin{equation}
\Box \varphi =\ast \mathrm{d}\ast \mathrm{d}\varphi =\frac{1}{\sqrt{g}}%
\partial _{\nu }\left[ \sqrt{g}g^{\nu \mu }\partial _{\mu }\varphi \right]
\label{thm}
\end{equation}
(cf. e.g. \cite{Nakahara} for a proof).
We can now immediately consider wave equations
in noncommutative spacetime. We just define the Laplace-Beltrami
operator by replacing the Hodge $\ast$-operator with the
$\ast^\FF$-operator,  for even dimensional noncommutative spaces with Lorenzian metric: $\square^\FF=\ast ^{\mathcal{F}}\mathrm{d}\ast ^{\mathcal{F
}}\mathrm{d}+ \mathrm{d}\ast ^{\mathcal{F
}}\mathrm{d}\ast ^{\mathcal{F}}$. In particular on a scalar field  we have
\begin{eqnarray} 
\label{scalar_eq}
  &&~~~\square ^{\mathcal{F}}\varphi =\ast ^{\mathcal{F}}\mathrm{d}\ast ^{\mathcal{F
}}\mathrm{d}\varphi =0~,
\end{eqnarray}
and more in general for a massive scalar field $(\square ^{\mathcal{F}}+m^2)\varphi=0$.

We can similarly consider Maxwell  equations on
noncommutative spacetime (without sources)
\begin{eqnarray}
&&
{{}^{{}^{\mbox{$\mathrm{d}{F} =0_{\phantom{M}}$}}}_{{}_{\mbox{$\mathrm{d}\ast ^{\mathcal{F}}{F} =0^{\phantom{M}}$}}}} 
\label{spin1_eq}~~~~~~~~~~~
\end{eqnarray}
The first equation implies (locally) the existence of a gauge potential 1-form
$A$ such that $F=dA$, the second one then becomes the noncommutative
Maxwell equation for the gauge potential $A$. These Maxwell equations
differ from the usual equations for a $U(1)$-gauge field in
noncommutative space, where $F=dA+A\wedge_\st A$. 

\section{Differential Geometry on  $\kappa$-Minkowski Spacetime\label{TPWM}}
The Weyl algebra or Poincar\'{e}-Weyl algebra
$\pw  =span\{M_{\mu \nu },P_{\mu ,}D\}$ is the 
extension of the Poincar\'e algebra with the dilatation generator $D$.
It is described by the following set of commutators and structure constants:
\begin{eqnarray}
\lbrack M_{\mu \nu },M_{\rho \lambda }] &=&i(\eta _{\mu \lambda }M_{\nu \rho
}-\eta _{\nu \lambda }M_{\mu \rho }+\eta _{\nu \rho }M_{\mu \lambda }-\eta
_{\mu \rho }M_{\nu \lambda }),  \label{MM} \\
\lbrack M_{\mu \nu },P_{\rho }] &=&i(\eta _{\nu \rho }P_{\mu }-\eta _{\mu
\rho }P_{\nu })\quad ,\qquad \lbrack P_{\mu },P_{\lambda }]=0,  \label{PP} \\
\left[ D,P_{\mu }\right] &=&iP_{\mu }\quad ,\quad \left[ D,M_{\mu \nu }%
\right] =0~, \label{DP}
\end{eqnarray}%
where $\eta_{\mu \nu }$ is the flat Minkowski space metric.
The representation of the Poincar\'e-Weyl generators as infinitesimal
transformations on Minkowski spacetime is given by the differential 
operators 
\begin{equation}
M_{\mu \nu }=i\left( x_{\mu
}\partial _{\nu }-x_{\nu }\partial _{\mu }\right)
\quad ;\quad  P_{\mu }=-i\partial _{\mu }
\quad ;\quad D=-ix^{\mu
}\partial _{\mu }~.  \label{diff_repr}
\end{equation}
The universal enveloping algebra of the Poincar\'e-Weyl algebra is
U\pw. This is a Hopf algebra with coproduct counit and antipode 
as recalled at the beginning of \S \ref{IIA}
(in particular $\Delta(u)=u\otimes 1+1\otimes u, \epsi(u)=0, S(u)=-u$
for all $u\in$ \pw).
\sk

We study a twist deformation of the Poincar\'e-Weyl algebra that leads 
to the noncommutative algebra of $\kappa$-Minkowski 
spacetime. 
We recall that the coordinates of $\kappa $-Minkowski
spacetime satisfy the commutation relations
\begin{equation}
 x^{0}\st x^{\/j}-x^{\/j}\st x^0=\frac{i}{\kappa }x^{\/j}\quad,
 \quad \quad x^{i}\st x^{j}-x^j\st x^i=0
\label{kmink}
\end{equation}
where $i$ and $j$ run over the space indices $1,\ldots n-1$.
For this twist deformation we list the corresponding 
quantum Lie algebra \pw$^\FF$, the quadratic Casimir
operator $\eta^{\mu\nu}P^\FF_\mu P_\nu^\FF$ of the Poincar\'e subalgebra, the 
commutation relations of the algebra of exterior forms, the
differential calculus and the (massless) fields wave equations, including
the Dirac equation. We show
that the d'Alembert opertor obtained from the quadratic Casimir coincides
with the Laplace-Beltrami operator obtained from the differential and the  Hodge $\ast^\FF$-operators. 

\subsection{The Jordanian twist}
The Jordanian twist 
of the Poincar\'e-Weyl algebra is defined by \cite{Borowiec:2008uj}: 
\begin{equation}
\mathcal{F}=\exp \left( -iD\otimes \sigma \right) \quad ;\quad \sigma =\ln
\left( 1+\frac{1}{\kappa }P_{0}\right) .  \label{nsymJ}
\end{equation}%
Its inverse is $\mathcal{F}^{-1}=\exp \left( iD\otimes \sigma \right) $. 
Equivalent expressions are
\eq
\mathcal{F}^{-1}=\sum_{n=0}^\infty \frac{1}{n!}(iD)^n\otimes \sigma^n
\;\quad \mbox{and}\quad  \FF^{-1}=\sum_{n=0}^\infty
\frac{(iD)^{\underline{n}}}{n!}\otimes \big(\frac{1}{\kappa}P_0\big)^n
\label{nsymJ2}
\en
where $X^{\underline{n}}=X(X-1)(X-2)\ldots (X-(n-1))$ is the so-called
lower factorial. The last
expression is the power series expansion in $\frac{1}{\kappa}$ of
$\FF^{-1}$ and  follows observing that $\FF^{-1}$ is analytic in $D$ and $P_0$
and that $D\otimes 1$ commutes with $1\otimes P_0$:
\begin{eqnarray}
\FF^{-1}&=&e^{(iD\otimes
  1)(1\otimes\sigma)}=(e^{1\otimes\sigma})^{iD\otimes
  1}=(e^{1\otimes\ln(1+\frac{1}{\kappa}P_0)})^{iD\otimes 1}
=(1\otimes (1+\frac{1}{\kappa}P_0))^{iD\otimes 1}\nonumber\\[.3em]
&=&
\sum_{n=0}^\infty\Big({}^{iD\otimes
  1_{\phantom{M}}}_{{\!^{\phantom{M}}~n}}\!\!\!\Big)(1\otimes
    \frac{1}{\kappa}P_0)^n=
\sum_{n=0}^\infty\Big({}^{iD_{\phantom{M}}}_{{\!\!\!^{\phantom{M}}n}}\!\!\!\Big)\otimes
    \big(\frac{1}{\kappa}P_0\big)^n=
\sum_{n=0}^\infty
\frac{(iD)^{\underline{n}}}{n!}\otimes \big(\frac{1}{\kappa}P_0\big)^n~,
\end{eqnarray}
 where
 $\Big({}^{X_{\phantom{M}}}_{{^{\phantom{M}}\!\!\!\!n}}\!\!\!\Big)=\frac{X^{\underline{n}}}{n!}$
   denotes the generalized binomial coefficient. 

\subsection{Quantum Poincar\'e-Weyl Lie algebra}\label{IIIA1}

The quantum Lie algebra \pw$^\FF$ has twisted generators
(cf. (\ref{twistedgen}), (\ref{DonUg})): 
\begin{equation}
P_{\mu }^{\mathcal{F}}=P_{\mu }\frac{1}{1+\frac{1}{\kappa }P_{0}}=P_{\mu
}e^{-\sigma }\quad ,\quad M_{\mu \nu }^{\mathcal{F}}=M_{\mu \nu }\quad
,\quad D^{\mathcal{F}}=D~.\label{FF40}
\end{equation}
In order to obtain the first expression use that
$\FF^{-1}=\exp(-iD\otimes -\sigma)$ and that $-iD$ on momenta acts as the identity operator:
$-iD(P_\mu)=[-iD,P_\mu]=P_\mu$, hence $\FF^{-1}=\exp(-iD\otimes
-\sigma)=\exp(1\otimes -\sigma)$ when we consider the adjoint action of
its first leg on $P_\mu$. Similarly, $M_{\mu\nu}^{\mathcal{F}}=M_{\mu\nu }$ and 
$D^{\mathcal{F}}=D$, because $D$ acts trivially on $M_{\mu\nu}$ and $D$.
The inverse of the universal $R$-matrix is
\eq
\RR^{-1}=\FF\FF_{21}^{-1}=e^{-iD\otimes\sigma}e^{\sigma\otimes
  iD}~.
\en
 In order to calculate the twisted commutators we first compute
\begin{eqnarray}
\RR^{-1}(P_\rho\otimes M_{\mu\nu})&=&P_\rho\otimes e^\sigma(M_{\mu\nu})=P_\rho\otimes M_{\mu\nu}+P_\rho\otimes
\frac{1}{\kappa}[P_0 , M_{\mu\nu}]\nonumber\\
\RR^{-1}(P_\mu\otimes D)&=&P_\mu\otimes e^\sigma(D)=P_\mu\otimes D-P_\mu\otimes \frac{i}{\kappa}P_0\label{Rm}
\end{eqnarray}
and $\RR^{-1} (M_{\rho\sigma}\otimes M_{\mu\nu})=M_{\rho\sigma}\otimes
  M_{\mu\nu}$, 
$\RR^{-1} (P_\nu\otimes P_\mu)=P_\nu\otimes P_\mu$, 
$\RR^{-1}(M_{\mu\nu}\otimes D)=M_{\mu\nu}\otimes D$.
From (\ref{Rm}) it immediately follows the nontriviality of the twisted
commutators (cf. (\ref{TwistedCom}))
\begin{eqnarray}
[M_{\mu\nu}^\FF,P_\rho^\FF]_\FF&=&M_{\mu\nu}^\FF P^\FF-P_\rho^\FF 
  M_{\mu\nu}^\FF-P_\rho^\FF\frac{1}{\kappa}([P_0,M_{\mu\nu}])^\FF~,\nonumber\\
{}[D^\FF,P_\mu^\FF]_\FF&=&D^\FF P^\FF_\mu-P^\FF_\mu D^\FF+P^\FF\frac{i}{\kappa}P_0~,
\end{eqnarray}
while the remaining twisted commutators are usual commutators. Use of
(\ref{FF40}), of the identities
$[M_{\mu\nu},e^{-\sigma}]=-e^{-2\sigma}
[M_{\mu\nu},e^\sigma]=-e^{-2\sigma}[M_{\mu\nu},\frac{1}{\kappa}P_0]$,
$[D,e^{-\sigma}]=-e^{-2\sigma}[D,e^\sigma]=-e^{-2\sigma}\frac{i}{\kappa}P_0$
and of the undeformed Poincar\'e-Weyl Lie algebra relations
then leads to the \pw$^\FF$ quantum Lie algebra 
\begin{eqnarray}\label{Jlie0}
\left[ M _{\mu \nu}^{\mathcal{F}},M_{\rho \lambda }^{\mathcal{F}}\right] _{
\mathcal{F}} &=&i(\eta_{\mu \lambda }M_{\nu \rho }^{\mathcal{F}}-\eta_{\nu
\lambda }M_{\mu \rho }^{\mathcal{F}}+\eta_{\nu \rho }M_{\mu \lambda }^{\mathcal{%
F}}-\eta_{\mu \rho }M_{\nu \lambda }^{\mathcal{F}}),  \label{Jlie1} \\
\left[ M_{\mu \nu }^{\mathcal{F}},P_{\rho }^{\mathcal{F}}\right] _{\mathcal{F%
}} &=&i(\eta_{\nu \rho }P_{\mu }^{\mathcal{F}}-\eta_{\mu \rho }P_{\nu }^{\mathcal{F%
}})\quad,\quad {\lbrack P_{\mu }^{\mathcal{F}},P_{\lambda }^{\mathcal{F}}]}_{\mathcal{F}}=0  \label{Jlie2} \\
\left[ D^{\mathcal{F}},P_{\lambda }^{\mathcal{F}}\right] _{\mathcal{F}}&=&iP_{\lambda }^{\mathcal{F}}, \quad ,\quad \left[ D^{\mathcal{F}},M_{\mu\nu}^{\mathcal{F}}\right] _{%
\mathcal{F}}=0~.  \label{Jlie3}
\end{eqnarray}
Notice that the structure constants are undeformed; this is also the
case for the Moyal-Weyl noncommutative Minkowski space \cite[\S 7.7]{book}, in
general however this does not hold, as the example of the twisted
quantum Lorentz Lie algebra 
in \cite[eq. (6.65)]{AC-SO} shows.

We complete the description of the quantum Lie algebra \pw$^\FF$  by listing
 the twisted coproduct and the antipode on the generators
\footnote{For sake of comparison we list also the twisted coproduct and antipode on the  untwisted generators (\ref{MM})-(\ref{DP}) because this is the usual set of generators considered for the Hopf algebra $U$\pw$^\FF$, even though they do not
  span a quantum Lie algebra as defined in {\it i), ii), iii)} before equation (\ref{adjact[]}).
\begin{eqnarray}
&&\Delta ^{\mathcal{F}}\left( P_{\mu }\right) =P_{\mu }\otimes e^{\sigma
}+1\otimes P_{\mu }, \\
&&\Delta ^{\mathcal{F}}\left( M_{\mu \nu }\right) =M_{\mu \nu }\otimes
1+1\otimes M_{\mu \nu }+\frac{1}{\kappa }D\otimes \left(\tau _{\mu
   }P_{\nu }
- \tau _{\nu }P_{\mu
}
\right) e^{-\sigma }, \\
&&\Delta ^{\mathcal{F}}\left( D\right) =D\otimes 1+1\otimes D-\frac{1}{%
\kappa }D\otimes P_{\tau }e^{-\sigma }=1\otimes D+D\otimes e^{-\sigma },
\end{eqnarray}%
\begin{eqnarray}
&&S^{\mathcal{F}}\left( P_{\mu }\right) =-P_{\mu }e^{-\sigma }, \\
&&S^{\mathcal{F}}\left( M_{\mu \nu }\right) =-M_{\mu \nu }+\frac{1}{\kappa }%
D\left(\tau _{\mu }P_{\nu }-\tau _{\nu }P_{\mu }\right) , \\
&&S^{\mathcal{F}}\left( D\right) =-D\big( 1+\frac{1}{\kappa }P_{\tau
}\big) =-De^{\sigma }~.
\end{eqnarray}
}
\begin{eqnarray}
\Delta ^{\mathcal{F}}\left( M_{\mu \nu }^{\mathcal{F}}\right) &=&M_{\mu \nu
}^{\mathcal{F}}\otimes 1+1\otimes M_{\mu \nu }^{\mathcal{F}}+\frac{1}{\kappa 
}D^{\mathcal{F}}\otimes \left( \tau _{\mu
}P_{\nu }^{\mathcal{F}}-\tau _{\nu }P_{\mu }^{\mathcal{F}}\right) \label{JtwcopM} \\
\Delta ^{\mathcal{F}}\left( P_{\mu }^{\mathcal{F}}\right) &=&P_{\mu }^{%
\mathcal{F}}\otimes 1+e^{-{\sigma}}\otimes P_{\mu }^{\mathcal{F}}
\label{JtwcopP} \\
\Delta ^{\mathcal{F}}\left( D^{\mathcal{F}}\right) &=& 1\otimes D^{\mathcal{F%
}}+D^{\mathcal{F}}\otimes e^{-{\sigma}}\label{JtwcopD} \\[1.2em]
S^{\mathcal{F}}\left( M_{\mu \nu }^{\mathcal{F}}\right) &=&-M_{\mu \nu }^{%
\mathcal{F}}+\frac{1}{\kappa }D^{\mathcal{F}}\left(\tau _{\mu }P_{\nu }^{\mathcal{F}}-\tau _{\nu }P_{\mu }^{\mathcal{F}}\right) e^{{\sigma}} \\
S^{\mathcal{F}}\left( P_{\mu }^{\mathcal{F}}\right) &=&-P_{\mu }^{%
\mathcal{F}}e^{{\sigma}}  \label{JtwSP} \\
S^{\mathcal{F}}\left( D^{\mathcal{F}}\right) &=&-D^{\mathcal{F}}e^{{%
\sigma}}\label{Jlielast}
\end{eqnarray}%
Where we have set $\tau ^{\mu }=(1,0,\ldots 0)$,
$\tau_\mu=\eta_{\mu\nu}\tau^\nu$, and $e^\sigma$ is here seen as dependent on  $P_0^\FF$ via
\begin{equation}
e^{{\sigma}} =1+\frac{1}{\kappa}P_0=\frac{1}{1-\frac{1}{\kappa }P_{0}^{\mathcal{F}}}~.
\label{tildeexp}
\end{equation}%
An easy way to
proceed in order to prove the twisted coproduct expressions
(\ref{JtwcopM})-(\ref{JtwcopD}) is to
recall the general one (\ref{eleven}) and use the properties
\begin{eqnarray}
\oR^\al\otimes\oR_\al(M_{\mu\nu})&=&\f^\al\otimes\f_\al(M_{\mu\nu})=1\otimes 
M_{\mu\nu}-iD\otimes \frac{1}{\kappa}[P_0,M_{\mu\nu}]\nonumber\\
\oR^\al\otimes\oR_\al(P_\mu)&=&\of_\al\otimes\of^\al(P_\mu)=e^{-\sigma}\otimes
P_\mu\nonumber\\
\oR^\al\otimes\oR_\al(D)&=&\f^\al\otimes\f_\al(D)=1\otimes 
D-D\otimes \frac{1}{\kappa}P_0~.
\end{eqnarray}
The twisted antipode on the generators  can then be easily obtained by applying
$S^\FF\otimes \id$ to (\ref{JtwcopD}), (\ref{JtwcopM}) and
$\id\otimes S^\FF$ to  (\ref{JtwcopP}) and then recalling the
Hopf algebra properties
$S^\FF(\xi_{1^\FF})\xi_{2^\FF}=\epsi(\xi)$ and $\xi_{1^\FF}S^\FF(\xi_{2^\FF}) =\epsi(\xi)$.

\sk

If we focus only on the algebra part of the quantum Lie algebra
\pw$^\FF$ we notice that  the bracket
$[~,~]_\FF$ closes on the twisted Poincar\'e generators. However, from the coproduct and the
antipode, because of the appearance of the dilatation generator, we see that the Poincar\'e generators do not form a quantum
Lie subalgebra of \pw$^\FF$. Notice however that the Lie algebra
$so(3)$ of
spatial rotations is an undeformed Lie subalgebra of the quantum Lie
algebra \pw$^\FF$ (indeed $M_{ij}^\FF=M_{ij},
\Delta^\FF(M_{ij}^\FF)=\Delta(M_{ij}), S^\FF(M_{ij}^\FF)=S(M_{ij})$). Moreover the twisted translations $P^\FF$
span a quantum Lie algebra that is a quantum Lie subalgebra of \pw$^\FF$ because the twisted bracket
${[~,~]}_\FF$, the deformed coproduct $\Delta^\FF$ and antipode $S^\FF$ on
translations are just expressed in terms of translations. Explicitly,
from (\ref{Jlie2}), (\ref%
{JtwcopP}) and (\ref{JtwSP}), the quantum Lie algebra of translations reads
\eq\label{quantumP}
{\lbrack P_{\mu
}^{\mathcal{F}},P_{\lambda}^{\mathcal{F}}]}_{\mathcal{F}}=0~,~~
\Delta ^{\mathcal{F}}\left( P_{\mu }^{\mathcal{F}}\right) =P_{\mu }^{%
\mathcal{F}}\otimes 1+e^{-{\sigma}}\otimes P_{\mu }^{\mathcal{F}}~,~~
S^{\mathcal{F}}\left( P_{\mu }^{\mathcal{F}}\right) =-P_{\mu }^{%
\mathcal{F}}e^{{\sigma}}  
\en
with $P_{\mu }^{\mathcal{F}}=\frac{P_\mu}{1+\frac{1}{\kappa}P_0}$ and $e^{{\sigma}} =1+\frac{1}{\kappa}P_0=\frac{1}{1-\frac{1}{\kappa }P_{0}^{\mathcal{F}}}$.
Thus the twisted momenta $\{P_{\mu }^\mathcal{F}\}$
generate a Hopf algebra: the quantum universal enveloping algebra of momenta.
All relevant formulae that in the following we derive for the
differential geometry on $\kappa$-Minkowski spacetime and the
dispersion relations depend only the quantum Lie algebra of
translations $P_{\mu }^{\mathcal{F}}$ summarized in (\ref{quantumP}). 

\subsubsection{Addition of momenta}
From the quantum Lie algebra of momenta we can immediately extract the
addition law of energy and momentum for multiparticle states. 
It is dictated by the coproduct (\ref{JtwcopD}) on the
quantum enveloping algebra.  In particular the total energy-momentum
${p_\mu^{\FF}\,}^{tot}$ of two free particles respectively of momenta
$p^\FF_\mu$ and $p'^\FF_\mu$, eigenvalues of the energy-momentum
operators $P_\mu^\FF$, is derived from (\ref{JtwcopP}) to be \begin{equation}\label{momenta+}
{p_\mu^{\FF\,}}^{tot} 
=p^\FF_\mu+p'^\FF_\mu -\frac{1}{\kappa}p^\FF_0p'^\FF_\mu~.
\end{equation}
We read from this expression a typical feature of addition of
momenta derived from quantum groups, cf. \cite{Lukierski:2002df},
the total momentum is not invariant under the exchange of the two
particles. This however does not mean that there is no symmetry
between the states $s$ and $s'$ of the unprimed and primed particle
(momentum eigenstates with eigenvalues $p^\FF$ and $p'^\FF$).  
The two particle states is the tensor product $s\otimes s'$. Instead of realizing the
permutation as $s\otimes s'\to s'\otimes s$, it is realized via the
appropriate action of the $R$-matrix as $s\otimes s'\to \oR^\al(s')\otimes \oR_\al(s)$;
hence the exchange of the two particles is implemented via the ${R}$-matrix.
  We remark that the present  ${R}$-matrix
$\RR=\FF_{21}\FF^{-1}$
is obtained from a twist 2-cocycle $\FF$ and hence it is
triangular and provides a representation of the
permutation group.

More importantly we notice that the deviation in (\ref{momenta+})
from the usual addition law is quadratic in the momenta (at all orders
in $\frac{1}{\kappa}$), and not
exponential like in $\kappa$-Poincar\'e in the standard basis
\cite{kpoin0} of generators as well as in the bicrossproduct basis
\cite{kpoin1}. 
Furthermore the total energy is invariant under the usual permutation of the two particles. These characteristics should
impose less stringent constraints on multiparticle applications of the
model. 

\subsubsection{Quadratic Casimir operator}
Associated with the quantum Lie algebra of momenta $\{P_{\mu
}^{\mathcal{F}}\}$ we have the quadratic Casimir operator
\begin{equation}
\square ^{\mathcal{F}}=P_{\mu }^{\mathcal{F}}P^{\mu \mathcal{F}}=P_{\mu
}P^{\mu }\frac{1}{\left( 1+\frac{1}{\kappa }P_{0}\right) ^{2}}=\square
e^{-2\sigma }  \label{def_casimir1}~.
\end{equation}%
The adjoint action of the twisted Poincar\'e generators on the
quadratic Casimir vanishes, and that of the dilatation generator is as in the
classical case: 
\begin{eqnarray}
\left[P_{\mu }^{\mathcal{F}}, \square ^{\mathcal{F}}\right] _{\mathcal{F}}
&=&0 \label{pdf}\\
\left[M_{\mu \nu }^{\mathcal{F}}, \square ^{\mathcal{F}},\right] _{\mathcal{F}%
} &=&0\label{mdf} \\
\left[D^{\mathcal{F}},\square ^{\mathcal{F}}\right] _{\mathcal{F}}
&=&-2i\square ^{\mathcal{F}}\label{ddf}
\end{eqnarray}%
where we have defined $[u^\FF,\xi^\FF]_\FF:=ad^\FF_{u^\FF\,}(\xi^\FF)=u^\FF_{~1^\FF}\xi^\FF S^\FF(u^\FF_{~2^\FF})$ for all 
$u\in g$ and $\xi\in Ug$. 
It is easy to prove the first relation: 
$\left[P_{\mu }^{\mathcal{F}}, \square ^{\mathcal{F}}\right] _{\mathcal{F}}=
P^\FF_\mu\square^\FF +e^{-\sigma}\square^\FF S^\FF(P_{\mu
}^{\mathcal{F}})=
P^\FF_\mu\square^\FF -e^{-\sigma}\square^\FF P_{\mu }^{\mathcal{F}}e^\sigma=0
$. The other relations can be similarly proven, although the required algebra
is longer. A quicker way is to use that the quantization map ${\cal D}$
intertwines the $U\pw$ and the $U\pw^\FF$ adjoint actions, see
\cite[Theorem 3.10]{AS}  (with $\mathbb{A} = U\pw$), so that
${\cal D}([ M_{\mu\nu}^\FF,\square])=[ M_{\mu\nu}^\FF,{\cal D}(\square)]_\FF$
and
${\cal D}([ D_{\mu\nu}^\FF,\square])=[ D_{\mu\nu}^\FF,{\cal D}(\square)]_\FF=-2i\square^\FF$.
Then from $[M_{\mu\nu}^\FF,\square]=[ M_{\mu\nu},\square]=0$ and $[
D^\FF_{\mu\nu},\square]=[D_{\mu\nu},\square]=-2i\square$ we
immediately obtain (\ref{mdf}) and (\ref{ddf}).

\sk

Instead of defining the quadratic Casimir  $\square^\FF=P_{\mu
}^{\mathcal{F}}P^{\mu \mathcal{F}}$ as the square (with the usual flat
Minkowski metric) of the twisted
momenta, one could deform the usual Casimir operator
using the quantization map (\ref{DonUg}). 
This deformation procedure leads to the same quadratic Casimir $\square^\FF$,
\begin{equation}\label{Dsquare}
{\cal D}(\square):=\of^\al(\square)\,\of_\al=\square e^{-2\sigma}=\square^\FF
\end{equation}
indeed $-iD(P_\mu)=P_\mu$ implies $-iD(\square)=2\square$ so that 
$\FF^{-1}=\exp \left( -iD\otimes -\sigma \right)=\exp \left( 2\otimes
  -\sigma \right)$ when we consider the adjoint action of its first leg
on $\square$.
In \S \ref{IIIA4} we further show that
$\square^\FF=\ast^\FF \d\ast^\FF\d$ when the quadratic Casimir is
represented as a differential operator on functions.

\sk
Finally we observe that the quadratic Casimir $\square^\FF$ 
coincides with the quadratic invariant considered by Magueijo and Smolin in \cite{MS1}. However the viewpoint on this
invariant is different: there momenta are undeformed while boosts are
deformed (and act
nonlinearly in momentum space),  here all Poincar\'e-Weyl
generators
are twist-deformed: momenta turn out to be nontrivially deformed
$P^\FF_\mu\not=P_\mu$  while boosts are trivially deformed
$M_{0j}^\FF=M_{0j}$. These twisted momenta  (rather than the
undeformed ones) are given physical relevance because they close a
quantum Lie algebra: that of translations in $\kappa$-Minkowski
noncommutative space. As we discuss in \S \ref{IVB} the physics of
the dispersion relations associated with the Casimir operator $\square^\FF$
will then turn out to be different from that considered in \cite{MS1}.

\subsection{$\kappa$-Minkowski spacetime and differential calculus}\label{IIIA2}
Using the representation (\ref{diff_repr}) of the Poincar\'e-Weyl 
generators as differential operators, the $\star $-product of two 
coordinate functions is easily seen to be 
$x^\mu\star x^\nu=\mu\{\FF^{-1}(x^\mu\otimes x^\nu)\}=x^\mu 
x^\nu-\frac{i}{\kappa}x^\mu\partial_0x^\nu$, henceforth the 
$\star$-commutator of the coordinates satisfies the 
$\kappa$-Minkowski relations (\ref{kmink}). 

The differential calculus on $\kappa$-Minkowski spacetime induced by
the Jordanian twist $\FF$ is easily described using the basis of 1-forms $\mathrm{d} x^\mu$.
The action of the dilatation and translation generators  is
given by the Lie derivative and on these forms it explicitly reads
\begin{equation}
D (\mathrm{d}x^{\mu })= \mathcal{L}_{-ix^\mu\partial_\mu}(\mathrm{d}x^{\mu
})=-i\mathrm{d}x^{\mu }~,
\quad P_\mu (\mathrm{d}x^{\mu })=\mathcal{L}_{-i\partial_{\mu }}(\mathrm{d}x^{\nu })=0~.  \label{lie_der}
\end{equation}%
Since in the Jordanian twist each term $\bar{\mathrm{f}}_{\alpha }$ in
the second leg of the tensor product $\mathcal{F}^{-1}=\bar{\mathrm{f}}^{\alpha }\otimes 
\bar{\mathrm{f}}_{\alpha }$ is a power of translation operators
 it is immediate to see that 
\begin{equation}
f\star \mathrm{d}x^{\mu } =f\mathrm{d}x^{\mu }  \notag ~,~~~f\st\mathrm{d}x^{\mu }\wedge _{\star }\mathrm{d}x^{\nu }=(f\st\mathrm{d}x^{\mu
})\wedge \mathrm{d}x^{\nu }=
f\mathrm{d}x^{\mu
}\wedge \mathrm{d}x^{\nu }
\end{equation}%
and iterating
\begin{equation}
f\st \mathrm{d}x^{\mu _{1}}\wedge _{\star }\mathrm{d}x^{\mu _{2}}\wedge _{\star
}\ldots \mathrm{d}x^{\mu _{r}}=f\mathrm{d}x^{\mu _{1}}\wedge \mathrm{d}x^{\mu
_{2}}\wedge \ldots \mathrm{d}x^{\mu _{r}}~.
\end{equation}%
The star product with functions on the right is however nontrivial
\begin{equation}
\mathrm{d}x^{\mu }\star f=\mathrm{d}x^{\mu }f-\frac{i}{\kappa }\mathrm{d}%
x^{\mu }\partial _{0} f =\mathrm{d}x^{\mu }(1-\frac{i}{\kappa }%
\partial _{0})f  \label{dxstf}
\end{equation}
so that 
\begin{equation}
f\star \mathrm{d}x^{\mu } -
\mathrm{d}x^{\mu }\star f
=\frac{i}{
\kappa}\partial_0 f \mathrm{d}x^{\mu } =\frac{i}{%
\kappa}\partial_0 f \star \mathrm{d}x^{\mu }~.  \label{relJf1form}
\end{equation}%
One can prove the first relation using the form of the twist (\ref{nsymJ})
written as a sum (recall that the twist acts via the Lie derivative on forms): 
\begin{eqnarray} 
\mathrm{d}x^{\mu }\star f&=&\sum_{n}\frac{1}{n!}\left(\LL_{{x^{\rho }\partial
_{\rho }}}\right) ^{\underline{n}}\left( \mathrm{d}x^{\mu }\right) \,\Big(\!-
\frac{i}{\kappa }\partial _{0}\Big) ^{n}\left( f\right)  \notag \\
&=&\mathrm{d}x^{\mu }f-\frac{i}{\kappa }\LL_{x^{\rho }\partial _{\rho} }\left( 
\mathrm{d}x^{\mu }\right)\, \partial _{0} f +\frac{1}{2!}\Big( 
\frac{i}{\kappa }\Big)^{2}\!\left(\LL_{x^{\rho }\partial _{\rho }}\right)^{
\underline{2}}\left( \mathrm{d}x^{\mu }\right) \,\partial _{0}^{2}f+...  \notag
\\
&=&\mathrm{d}x^{\mu }f-\frac{i}{\kappa }\mathrm{d}x^{\mu }\partial
_{0}f
\end{eqnarray}
where we used the definition of the lower factorial, i.e. $X^{\underline{n}%
}=X(X-1)...(X-n+1)$, and the fact that $\left(\LL_{x^{\rho }\partial _{\rho
}}\right)^{\underline{2}}\left( \mathrm{d}x^{\mu }\right) =\left(\LL_{x^{\sigma
}\partial _{\sigma }}-1\right)  \LL_{x^{\rho }\partial _{\rho }}
\left( \mathrm{d}x^{\mu }\right) =
\left(\LL_{x^{\sigma
}\partial _{\sigma }}-1\right) 
\mathrm{d} \!\left( \LL_{x^{\rho }\partial _{\rho }} x^{\mu }\right) =
\left( \LL_{x^{\sigma }\partial _{\sigma
}}-1\right)  \mathrm{d}x^{\mu }=0$, so that also all the higher order terms vanish.
Expression (\ref{relJf1form}) for $f=x^\nu$ has been considered also in \cite{Mel} (as
special case ${\cal S}_1$ in their equation (18)).

{}From (\ref{relJf1form}) it follows that
\begin{equation}
f\star \mathrm{d}x^{\mu }=\mathrm{d}x^{\mu }\star \frac{1}{(1-\frac{i}{
\kappa }\partial_{0})}f=\mathrm{d}x^{\mu }\star e^{-\sigma }(f)  \label{comm_f1formxx}~.
\end{equation}
Iterating this expression
and using associativity of the $\wedge_{\star }$-product we immediately obtain: 
\begin{equation}
f\star \mathrm{d}x^{\mu _{1}}\wedge _{\star }\mathrm{d}x^{\mu
_{2}}...\wedge _{\star }\mathrm{d}x^{\mu _{r}}=\mathrm{d}x^{\mu
_{1}}\wedge _{\star }\mathrm{d}x^{\mu _{2}}...\wedge _{\star }\mathrm{d}%
x^{\mu _{r}}\star \frac{1}{\left( 1-\frac{i}{\kappa }\partial _{0}\right)
^{r}}f~.\label{75}
\end{equation}
\sk
We now express the exterior derivative on $\kappa$-Minkowski spacetime
in terms of the  twisted momenta $P^\FF_\mu$, thus confirming that they
have the interpretation of quantum infinitesimal transformations.
Corresponding to the representation $P_\mu=-i\partial_\mu$ and the
relation $P^\FF_\mu =P_{\mu }\frac{1}{1+\frac{1}{\kappa }P_{0}}$ 
we have the representation $P^\FF_\mu=-i\partial^\FF_\mu$, where 
\eq
\partial^\FF_\mu:=\frac{1}{1-\frac{i}{\kappa }\partial_0} \partial _{\mu }~.
\en
The twisted momenta act as quantum infinitesimal translations because for
any function $f$ on $\kappa$-Minkowski spacetime we have
\eq 
\d f= \d x^\mu\star iP^\FF_\mu(f)~,\label{PFdiff}
\en
i.e., $\d f=\d x^\mu\star \partial^\FF_\mu f$.
Indeed, $\d x^\mu\star\partial^\FF_\mu f=
\d x^\mu\star\frac{1}{1-\frac{i}{\kappa}\partial_0} \partial _{\mu
}f= \d x^\mu\,\partial _{\mu }f=\d f$, where we used (\ref{dxstf}) and
that the exterior derivative is the usual undeformed one (cf. \S \ref{DiffCalc}).
It is also instructive to see that the deformed Leibniz rule
(\ref{LeibFF}), that in the present representation reads, cf. (\ref{JtwcopP}),
\eq
\partial^\FF_\mu (f\st h)=\partial^\FF_\mu
(f)\st h+e^{-\sigma}(f)\st \partial^\FF_\mu (h)~,
\en
combines with the commutation property (\ref{comm_f1formxx}) to 
give the undeformed Leibniz rule for the exterior derivative
$\d (f\st h)=\d f\st h +f\st \d h$.

\subsection{Hodge star operator}\label{IIIA3}
The Hodge $\ast^\FF$-operator is defined in (\ref{quantum_ast}) as the quantization $\ast^\FF={\cal
  D}(\ast)$  of the Hodge $\ast$-operator.
On an $s$-form $\d x^{\mu _{1\!}}\wedge_\st\d x^{\mu _{2\!}} ...\wedge_\st \d x^{\mu _{s}}=\d x^{\mu _{1\!}}\wedge\d x^{\mu _{2\!}} ...\wedge \d x^{\mu _{s}}$,
it  reads (in the last line recall that
each term $\bar{\mathrm{f}}_{\alpha }$ in
the second leg of the tensor product $\mathcal{F}^{-1}=\bar{\mathrm{f}}^{\alpha }\otimes 
\bar{\mathrm{f}}_{\alpha }$ is a power of translation operators,
 so that it acts trivially on $\d x^{\mu _{1}}\!\wedge_\st \d x^{\mu_{2}}... \wedge_\st \d x^{\mu _{s}}$)
\begin{eqnarray}
\ast ^{\mathcal{F}}\left( \d x^{\mu _{1}}\!\wedge_\st \d x^{\mu
    _{2}}... \wedge_\st \d x^{\mu _{s}}\right)
&=&\mathrm{\bar{f}}_{1}^{\alpha }\Big(\ast \big( S\left( \mathrm{%
\bar{f}}_{ 2 }^{\alpha }\right) \mathrm{\bar{f}}_{\alpha
}\left( \d x^{\mu _{1}} \!\wedge_\st\d x^{\mu _{2}} ... \wedge_\st \d
  x^{\mu _{s}}\right) \!\big)\!\Big)\nonumber\\[.2em]
&=&\ast \left(
\d x^{\mu _{1}}\!\wedge_{\st}  \d x^{\mu _{2}} ... \wedge_{\st} \d x^{\mu _{s}}\right)  \label{J_quantum_hodge}~.
\end{eqnarray}
Hence the Hodge
$\ast^\FF$-operator equals the commutative Hodge $\ast$-operator on the exterior forms $\d x^{\mu _{1}}\!\wedge_\st \d x^{\mu_{2}}... \wedge_\st \d x^{\mu _{s}}$.
Furthermore, recalling that the Hodge star $\star ^{\mathcal{F}}$ is right
$A_\st$-linear we have the general expression:
\begin{equation}
\ast ^{\mathcal{F}}\left( \d x^{\mu _{1}}\wedge_\st \d x^{\mu _{2}} ...\wedge_\st \d x^{\mu 
    _{s}}\star f\right)=\ast \left(
\d x^{\mu _{1}}\wedge_\st \d x^{\mu _{2}}....\wedge_\st  \d x^{\mu 
  _{s}}\right) \star f\label{77}
~;
\end{equation}
explicitly, $\ast ^{\mathcal{F}}\left( \d x^{\mu _{1}}\wedge_\st \d x^{\mu _{2}} ...\wedge_\st \d x^{\mu
    _{s}}\star f\right)=\frac{1}{(n-s)!}\varepsilon^{\mu_1\mu_2...\mu_s}{}_{\nu_{s+1}\nu_{s+2}...\nu_n}\d
x^{\nu_{s+1} }\wedge_\st \d x^{\nu_{s+2} } ...\wedge_\st \d x^{\nu_{n}
}\star f$.
In particular we see that $\ast^\FF$ squares to  $\pm $ the  
identity. 

 It is now easy to show that for an arbitrary form $\omega\in
 \Omega_\st^s$ of homogeneous
degree $s$ in $n$-dimensional $\kappa$-Minkowski space we have
$\ast^\FF(\omega)=(1-\frac{i}{\kappa}\partial_0)^{n-2s}\,\ast(\omega)$. Indeed, use of
(\ref{75}) and (\ref{77}) gives
\begin{eqnarray}
\ast^\FF(f\star  \d x^{\mu _{1}}\wedge_\st \d x^{\mu _{2}} ...
\wedge_\st \d x^{\mu_{s}}) &=&\ast \big(\d x^{\mu _{1}}\wedge_\st \d x^{\mu _{2}} ... 
\wedge_\st \d
                               x^{\mu_{s}}\big)\st\big(1-\frac{i}{\kappa}\partial_0\big)^{-s}f\nonumber\\
&=&\big(1-\frac{i}{\kappa}\partial_0\big)^{n-2s}f\st\big(\ast \!\big(\d x^{\mu _{1}}\wedge_\st \d x^{\mu _{2}} . . . 
\wedge_\st \d x^{\mu_{s}}\big)\big)\nonumber\\
&=&\big(1-\frac{i}{\kappa}\partial_0\big)^{n-2s}\,\ast \!\big(f\st\d x^{\mu _{1}}\wedge_\st \d x^{\mu _{2}} . . . 
\wedge_\st \d x^{\mu_{s}}\big)~.\label{80ast}
\end{eqnarray}

\subsection{Field equations}\label{IIIA4}

\paragraph{Scalar fields}
The d'Alembert operator $\square^\FF$ on
$\kappa$-Minkowski spacetime can be defined:\begin{itemize}\item[(1)] as the quadratic Casimir
$P_\mu^\FF P^{\mu\FF}$ (once the generators of translations are
represented as differential operators);
\item[(2)] as the quantization of the
 d'Alembert operator on commutative Minkowski spacetime ${\cal D}(\square):=
\mathrm{\bar{f}}^\alpha(\square)\mathrm{\bar{f}}_{\alpha}$,
cf. equation (\ref{Dsquare});

\item[(3)] via the Hodge
$\ast^\FF$-operator, as the Laplace-Beltrami operator $\ast^\FF
\d\ast^\FF \d\,$.
\end{itemize}
These definitions coincide: We already saw in (\ref{Dsquare}) the
  equivalence of the first two definitions, for the third definition, use of (\ref{80ast}) gives
$\ast^\FF \d\ast^\FF \d \varphi= \frac{1}{(1-\frac{i}{%
\kappa }\partial_{0})^2}\square \varphi$ showing the equality with
the other two.
In conclusion for a (massless) scalar field $\varphi$ we have the wave equation
\begin{equation}  \label{J_scalar_eq}
\square ^{\mathcal{F}}\varphi =\ast^\FF
\d\ast^\FF \d \varphi=\frac{1}{(1-\frac{i}{%
\kappa }\partial_{0})^2}\square \varphi =0~,
\end{equation}
where $\square=\partial_\mu\partial^\mu$.

 Notice that this wave
equation is equivalent to the undeformed one $\square \varphi =0$,
indeed the differential operator $\frac{1}{(1-\frac{i}{%
\kappa }\partial_{0})^2}$ is invertible. This result agrees with the
well known one for free fields in noncommutative Moyal-Weyl space.
There the result is trivial because the $\star$-product is simpler and
translation invariant  (so that $P_\mu{^{\FF_{MW}}}=P_\mu$ where
$\FF_{MW}=e^{-i\theta^{\mu\nu}\partial_\mu\otimes\partial_\nu}$). 
The Moyal-Weyl $\st$-product is also compatible with the usual
integral on commutative Minkowski space, thus already the free field
actions coincide $\int \partial_\mu\varphi\star\partial^\mu\varphi\,
\d^nx=\int \partial_\mu\varphi\partial^\mu\varphi\, \d^nx$. 
Notice however that while in Minkowski spacetime with Moyal-Weyl
noncommutativity the equations of motion and the action remain
undeformed also when considering massive scalar fields, $\int \partial_\mu\varphi\star\partial^\mu\varphi-m^2\varphi\star\varphi\,
\d^nx=\int \partial_\mu\varphi\partial^\mu\varphi-m^2\varphi^2\, \d^nx$, this is no more the case in
$k$-Minkowski spacetime, indeed 
\begin{equation}\label{weqm}
(\square ^{\mathcal{F}} +m^2)\varphi=0
\end{equation}
is not equivalent to $(\square+m^2)\varphi=0$. It is easy to check
covariance of (\ref{weqm}) under the quantum Lie algebra of momenta $P^\FF_\mu$, indeed
\eq
P^\FF_\mu\big((\square ^{\mathcal{F}} +m^2)\varphi\big)=
(\square ^{\mathcal{F}} +m^2) P^\FF_\mu\varphi\en
as follows from considering the coproduct (\ref{JtwcopP}) and recalling the
triviality of the quantum adjoint action of $P^\FF_\mu$ on $\square^\FF$,
cf. equation (\ref{pdf}).
\sk 

Among the three considered formulations of the operator $\square^\FF$, the advantage of the Laplace-Beltrami
operator formulation is that it immediately applies to the case of curved
and noncommutative spacetime. As shown in \cite{ABP2}, when a
gravitational background is turned on then the wave equation in
noncommutative spacetime in general differs from that in commutative spacetime.

\sk
\paragraph{Spin 1 fields and twisted Maxwell equations}
Maxwell equations in $\kappa$-Minkowski spacetime read
$\mathrm{d}{F}=0$, $\d\ast^\FF \!F=0$. The first equation is undeformed
because the exterior derivative is undeformed. Also the second equation is
equivalent to the undeformed one $\d\ast F=0$, indeed 
$\d\ast^\FF \!F=\d (1-\frac{i}{%
\kappa }\partial_{0})^{n-4}\ast F=(1-\frac{i}{%
\kappa }\partial_{0})^{n-4}\d \ast F$, and since $(1-\frac{i}{%
\kappa }\partial_{0})^{n-4}$ is invertible  $\d\ast^\FF \!F=0$ is
equivalent to  $\d\ast F=0$. This is no more the case if sources are
present or if we are in curved noncommutative spacetime.
\sk

\paragraph{Spin 1/2 fields and twisted Dirac equation}

The classical Dirac operator in Minkowski spacetime,
$
{\slash\!\!\!\partial}
$,
can be written as $-\gamma^\mu P_\mu$ and is twisted according to
the quantization map ${\cal D}$ by considering the twisted momenta ${\cal D}(P_\mu)=P^\FF_\mu$,
so that we have $-\gamma^\mu
P^\FF_\mu$, i.e., 
\begin{equation}
{\slash\!\!\!\partial}^\FF=i\gamma ^{\mu }\partial _{\mu }^{\mathcal{F}%
}=i\gamma ^{\mu }\frac{1}{1-\frac{i}{\kappa }\partial_0} \partial _{\mu }=\frac{1}{1-\frac{i}{\kappa }\partial_0} {\slash\!\!\!\partial}~.
\end{equation}
It is immediate to check that the twisted Dirac operator squares
to the twisted d'Alembert opertor
${({\slash\!\!\!\partial}^{\mathcal{F}})}^2=\Box ^{\mathcal{F}}$.
The Dirac equation for a massless spinor field $\psi$ reads
\begin{equation}
{\slash\!\!\!\partial}^\FF\psi=0~;
\end{equation}
it is equivalent to the undeformed one because  
$
\frac{1}{1-\frac{i}{\kappa }\partial_0} 
$ is invertible.
As in the case of scalar fields, the massive Dirac field
equation on $\kappa$-Minkowski however differs from the undeformed one.

\section{Dispersion Relations in $\kappa$-Minowski Spacetime}\label{IV}

Given the Jordanian twist $\FF$ in (\ref{nsymJ}), we derived the wave
operator $\square^\FF$ for scalar fields
in flat noncommutative spacetime using three different techniques that
lead to the same result: $\square^\FF$  as the quadratic Casimir 
$P_\mu^\FF {P^{\mu}}^\FF$, as the deformation ${\cal D}(\square)$ of
the commutative wave operator $\square$, and  as the
Laplace-Beltrami operator $\square^\FF=\ast^\FF \d\ast^\FF \d$. 

We are now in a position to study the relation between energy, momentum,
frequency and wave vector and the corresponding dispersion
relations. We first study the massless case 
considering both the momentum space and the position space perspectives.
We similarly study the massive case. There, as discussed in the
introduction, we can consider masses as emerging from a conformal anomaly
and the $\kappa$-Poincar\'e-Weyl symmetry as a broken symmetry. 

\subsection{Massless fields:  
undeformed dispersion relations and modified Einstein-Planck relations} 
The  massless wave equation is undeformed, indeed 
$\square^\FF\varphi=0$ is equivalent to $\square \varphi=0$. 
The {\it energy - momentum dispersion relations} are $P_\mu^\FF
{P^{\mu}}^\FF=0$, and are undeformed as 
well. We immediately have (inserting $c$ and setting
${P^\FF\,}^2=P_i^\FF {P^{i\,\FF}}$) $\frac{\d  E^\FF}{\d P^\FF}=c$. 
More elegantly, let $\varphi(\xb,t)$ be an eigenvector of  the
energy - momentum operators (cf. (\ref{diff_repr}) and (\ref{FF40})) 
\eq\label{PFdiffop}
P^\FF_\mu=\frac{P_\mu}{1+\frac{1}{\kappa}P_0}=
\frac{-i\partial_\mu}{1-\frac{i}{\kappa}\partial_0}
\en
with eigenvalues
  $p_\mu^\FF=(-E^\FF, \pb^\FF)$, 
i.e.,  ${p^\mu}^\FF=(E^\FF, \pb^\FF)$,
then $\varphi(\xb,t)$ is a solution of
  the wave equation $\square^\FF\varphi=0$  if $p_\mu^\FF
{p^{\mu}}^\FF=0$, hence  $\frac{\d  E^\FF}{\d p^\FF}=c$ (with $p^\FF$
the modulus of $\pb^\FF$).

\sk
The same result follows if we consider the {\it frequency -- wave vector
dispersion relations}. To this aim we observe that  the wave functions
$\varphi(\xb,t)$ of definite momentum are proportional to
$e^{i(\kb\xb-\omega t)}$, and that  they are a solution of the
wave
equation $\square^\FF\varphi=0$ if the usual undeformed
dispersion relation holds, $\om^2=c^2 k^2$. The 
group velocity is therefore undeformed as well,
$
{v}_g=\frac{\d \om}{\d k}=c$, and $v_g=\frac{\d  E^\FF}{\d p^\FF}
$.

Finally, evaluation of the energy momentum operator (\ref{PFdiffop}) 
on the monochromatic wave $e^{i(\kb\xb-\omega t)}$ leads to
the modified Einstein-Planck 
relations
\eq
E^\FF=\frac{\hbar \omega}{1-\frac{\omega}{c\:\!\kappa}}~,~~
\pb^\FF=\frac{\hbar \kb}{1-\frac{\omega}{c\:\!\kappa}}~\label{eigvl}~.
\en 
If we assume a negative value for the deformation parameter $\kappa$
then we can define $E_{p}=- c\:\!\hbar\kappa$ that we identify with Planck
energy. In this case we have 
\eq
E^\FF=\frac{\hbar \omega}{1+\frac{\hbar\omega}{E_p}}~,~~
\pb^\FF=\frac{\hbar \kb}{1+\frac{\hbar\omega}{E_p}}~\label{eigvlEp}~;
\en 
these deformed Einstein-Planck relations imply a maximum energy reachable by an
elementary particle: Planck energy. 
The other choice $E_{p}=c\:\!\hbar\kappa$, corresponding to a positive
value of the deformation parameter $\kappa$ is also possible, in this
case we have a maximum frequency $c\kappa$ that is associated with
infinite energy.
\sk
In conclusion, for massless elementary particles we have the usual
dispersion relations, however the Einstein-Planck relations
are deformed so that there is a maximum energy reachable by an
elementary particle: Planck energy, or, depending on the sign of the
deformation parameter, a maximum frequency.
\sk
\subsection{Massive fields}

We proceed similarly with the equation for massive fields.
An eigenvector $\varphi(\xb, t)$ of the energy-momentum operators
$P_\mu^\FF$ with eigenvalues $p_\mu^\FF=(-E^\FF, \pb^\FF)$ solves the wave
equation  $(\square^\FF+ m^2)\varphi=0$ if the usual energy-momentum dispersion relations
$(E^\FF)^2=(p^\FF)^2+m^2$ hold. Hence, inserting $c$,
$\frac{\d  E^\FF}{\d p^\FF}=c^2\frac{p^\FF}{E^\FF}$.
The Einstein-Planck and de
Broglie relations are as in (\ref{eigvl}).  Therefore for negative
deformation parameter we can set $E_p=-c\hbar\kappa$ and we see that
also massive elementary
particles have Planck energy as maximum energy, and correspondingly a
maximum momentum. On the other hand, for positive valued deformation
parameter we can set $E_p=c\:\!\hbar\kappa$ and we have the maximum
frequency $c\:\!\kappa$.

The frequency-wave vector dispersion relations are modified to 
\eq
\omega^2-c^2 k^2=\frac{m^2c^4}{\hbar^2}\left(1-\frac{\omega}{c\:\!\kappa}\right)^2\,;
\en
and therefore they differ from the energy -- momentum dispersion
relations.
In particular the group velocity at first order in $\Ok$ reads 
$$
v_g=\frac{\d \om}{\d k}=c^2\frac{k}{\om}(1+\frac{m^2c^3}{\hbar^2\om \kappa})+\OK~
$$
and differs from $\frac{\d  E^\FF}{\d p^\FF}=c^2\frac{p^\FF}{E^\FF}=c^2\frac{k}{\om}$ (cf. (\ref{eigvl})).
It is easy to see from $(E^\FF)^2=(p^\FF)^2+m^2c^4$ and (\ref{eigvl})
that at first order in the deformation this difference is limited by
the ratio $mc^2/E_p$.  The discrepancy between $v_g$ and $\frac{\d
  E^\FF}{\d p^\FF}$ is thus proportional to the particle mass and
inversely proportional to $E_p$ or $\kappa$. These are indeed the two
scales arising in the  breaking of  conformal
symmetry scenario discussed
in the introduction. 

\subsection{Relation to Deformed Special Relativity (DSR) theories}\label{IVB}
It is very interesting to compare the energy-momentum undeformed dispersion relations result,
that we have derived from a general noncommutative differential
geometry construction, with previous results in the literature, in particular with the Deformed Special Relativity (DSR)
proposed in \cite{MS1}, \cite{MS2}, also known as DSR2.
There, Magueijo and Smolin consider a nonlinear modification of the
action of the Lorentz generators on  momentum space. It is induced by
the differential operator $U=e^{{\la p_0 p_\mu\frac{\partial}{\partial p_\mu}}}$
on momentum space itself. Since it is the exponential of a vector
field it is an element, say $\varphi^{_{U\!}}$, of the group of
diffeomorphism on momentum space. For $f,h$ functions on momentum
space we then have 
$U(f)(p)=f(\varphi^{_U\!}(p))$, $U(h)(p)=h(\varphi^{_U\!}(p))$, and from
$(fh)(\varphi^{_{U}\!}(p))=f(\varphi^{_U\!}(p))\,
h(\varphi^{_U\!}(p))$ we immediately obtain  $U(fh)=
U(f)U(h)$.
Hence the action of $U$ is determined by that on the coordinate functions,
that explicitly reads
\begin{equation}
p_\mu\to U(p_\mu)= \frac{p_\mu}{1-\la p_0} ~.
\end{equation}
The Lorentz generators $M_{\mu}^{~\nu}=p_{\mu}\frac{\partial}{\partial
  p_\nu}-p_{\nu}\frac{\partial}{\partial p_\mu}$ are correspondingly transformed to
$M_{\mu}^{~\nu}\to U M_{\mu}^{~\nu}U^{-1}$  (cf. the adjoint action
$ad_U(M_{\mu}^{~\nu})$ in (\ref{adact})), explicitly $M_{i}^{~j}\to M_{i}^{~j}$ and 
$M_{0}^{~i}\to M_{0}^{~i}+\la p_ip_\mu\frac{\partial}{\partial p_\mu}$
[hint: use $U\frac{\partial}{\partial p_\mu} U^{-1}=(1-\la
p_0)(\frac{\partial}{\partial p_\mu}-\delta_{\mu 0}\la p_\nu \frac{\partial}{\partial p_\nu})$].
Under the modified action of the Lorentz generators the usual
quadratic expression ${\eta^{\mu\nu}p_\mu p_\nu}$ is no more
invariant, the new invariant is 
\begin{equation}\label{SMinvariant}
U(\eta^{\mu\nu}p_\mu p_\nu)=\frac{\eta^{\mu\nu}p_\mu p_\nu}{(1-\la p_0)^2}~.
\end{equation}
Upon the identification $\frac{1}{\kappa}=-\la$, and observing that
the translation generators $P_\mu$ act in momentum space via
multiplication by $p_\mu$, we see from equation (\ref{FF40})  that the
twist quantization map on momenta
\begin{equation}
P_\mu\to   P_\mu^\FF={\cal D}(P_\mu)=\frac{P_\mu}{1+\frac{1}{\kappa} P_0} 
\end{equation}
equals the $U$ transformation map.
This holds more in general for functions of momenta, indeed
$(P_{\mu_1}P_{\mu_2}\ldots P_{\mu_n})^\FF:={\cal D}(P_{\mu_1}P_{\mu_2}\ldots P_{\mu_n})=
P_{\mu_1}^\FF P_{\mu_2}^\FF\ldots P_{\mu_n}^\FF$ (for a proof just recall the
derivation of  (\ref{Dsquare})), so that $(fh)^\FF=f^\FF h^\FF$ for
$f$ and $h$ functions of  $P_\mu$, i.e. we recover the algebra
property of $U$: $U(fh)= U(f)U(h)$.

We have shown that the quantization map ${\!\!\phantom I}^\FF={\cal D}$
corresponding to the Jordanian twist (\ref{nsymJ}) equals, on momentum space,  the $U$ transformation introduced in \cite{MS1}. However the action of ${\cal D}$ is independent from the
representation used (${\cal D}$ acts on the Poincar\'e-Weyl generators
irrespectively of their representation in momentum or position space),
therefore the present result allows to extend the construction in DSR2
from operators in momentum space to operators in position space. 

In DSR2 the change to the new invariant $\eta^{\mu\nu}p_\mu p_\nu\to 
\frac{\eta^{\mu\nu}p_\mu p_\nu}{(1-\la p_0)^2}$
is interpreted as giving rise to modified dispersion relations because
the physical momenta are considered to be the usual undeformed ones,
see also \cite{Visser, Lukierski:2002df}.
Otherwise stated, the transformation $Q\to U Q
U^{-1}$ applies only to the Lorentz generators, it does not apply to
the whole Poincar\'e generators\footnote{
Since $P_\mu$ on momentum space acts
simply as multiplication by $p_\mu$, the transformation $Q\to U Q
U^{-1}$ on momenta reads
$P_\mu\to UP_\mu U^{-1}= \frac{P_\mu}{1-\la P_0}$, indeed 
$UP_\mu U^{-1}(f)=U(p_{\mu\,} U^{-1}(f))=U(p_\mu)f=\frac{p_\mu}{1-\la
  p_0}f=
\frac{P_\mu}{1-\la P_0}(f).$}.
In the present paper we similarly consider the change to the new invariant
$\eta^{\mu\nu}p_\mu p_\nu\to 
\frac{\eta^{\mu\nu}p_\mu p_\nu}{(1-\la p_0)^2}$, it is written as $\square\to \square^\FF$ 
in (\ref{def_casimir1}), (\ref{Dsquare}); and $\square^\FF$ is proven
to be an invariant under
quantum Poincar\'e transformations in (\ref{pdf}), (\ref{mdf}).
However we follow a different route from that of DSR theories. We have first
singled out the quantum Lie algebra (\ref{Jlie0})-(\ref{Jlielast}) of
the quantum group of Poincar\'e-Weyl (i.e., of the twist deformed universal
enveloping algebra of  Poincar\'e-Weyl). This is generated by $P^\FF,
M^\FF_{\mu\nu}, D^\FF$. Then we have shown in (\ref{PFdiff}) that the
differential calculus on $\kappa$-Minkowski space is defined in terms
of the generators $P^\FF$, that  hence have not only the Lie-algebraic
but also the differential geometry meaning of generators of translations. We therefore conclude that these are the generators
of the Poincar\'e-Weyl group that encode the physics of energy and momentum.
As shown at the beginning of this section, in terms of these momenta
the dispersion relations for massless and massive particles are
undeformed. We hence arrive at different conclusions with respect to
\cite{MS1}, where energy-momentum dispersion relations for massless
particles are undeformed, but for massive particles are deformed.

\section{Conclusions}
Deformed Special Relativity theories have been considered as
phenomenological models providing an effective description of quantum
spacetime. The associated wave equations can be obtained as wave
equations arising in $\kappa$-Minkowski spacetime. 
The relation between noncommutative spacetimes, dispersion
relations and DSR theories depends however on the different choices
and realization of the algebra of momenta and coordinates in
noncommutative space. This issue has emerged in
particular considering  $\kappa$-Minkowski as homogeneous space under
the $\kappa$-Poincar\'e quantum symmetry group. There different  nonlinearly related sets of
generators of the $\kappa$-Poincar\'e algebra lead to different
dispersion relations \cite{KG2002we} as well as
different realizations of the $\kappa$-Minkowski coordinates
 \cite{BGMP}, see also \cite{SIGMA} and \cite{Mel}. We resolved the
 ambiguities associated with the choice of the basis of momenta and
 coordinates and their realization by singling out the quantum
 Lie algebra, eq.  (\ref{twistedgen}), of the given quantum universal enveloping
 algebra  (quantum symmetry group), and by
constructing a coordinate independent noncommutative differential
geometry (build on the exterior derivative, the $\ast^\FF$ Hodge star
operator, the twisted d'Alembertian $\square^\FF=\ast^\FF\d
\ast^\FF\d$).  
We exemplified the construction in the case of the Jordanian twist deformed
Poincar\'e-Weyl group with $\kappa$-Minkowski
spacetime as its homogeneous space. While the twist depends on the
dilatation generator, all relevant formulae for the
differential geometry on $\kappa$-Minkowski spacetime and the
dispersion relations depend only the quantum Lie algebra of
translations $P_{\mu }^{\mathcal{F}}$, summarized in (\ref{quantumP}). The quantum Lie algebra
construction can in principle be carried out also for the
$\kappa$-Poincar\'e group, however it is not canonically determined as
in the case of quantum groups obtained via twists. Strikingly, the example considered leads to
the wave equation studied in DSR2, hence extending it from
momentum space to position space.  
It would be interesting to further investigate DSR theories and the
associated relative locality principle \cite{ACFKGS}, including the
corresponding interpretation of the dispersion relations (see \cite{AmelinoCamelia:2011cv} for a critical discussion), with the perspectives and techniques provided in this paper.

\section*{Acknowledgements}
The authors would like to thank L.$\,$Castellani
and  J.$\,$Lukierski for useful discussions. This work begun during a three month visit of A.$\,$B. at
Universit\`a del Piemonte Orientale supported by the university funds
for internationalization.
P.$\,$A. is partially supported by INFN, CSN4, iniziativa specifica
GSS
and is affiliated to INdAM, GNFM (Istituto Nazionale di Alta Matematica, Gruppo Nazionale di Fisica Matematica).
A.$\,$B is supported by NCN (Polish National Science Center) project
2014/13/B/ST2/04043.
P.$\,$A. and A.$\,$B. acknowledge also the support from COST (European Cooperation in
Science and Technology) Action MP1405 QSPACE.
A.$\,$P. acknowledges the funding from the European Union's Horizon 2020 research
and innovation programme under the Marie Sklodowska-Curie grant agreement No
660061.

\end{document}